\documentclass[final,3p,times,twocolumn]{elsarticle}

\usepackage{amssymb}
\usepackage{amsmath}

\usepackage{lineno}

\usepackage{blindtext}
\usepackage{cuted}
\usepackage{hyperref}
\RequirePackage[capitalise]{cleveref}
\usepackage{mymacros}
\usepackage{animate}
\SetKwComment{Comment}{/* }{*/}

\usepackage{tablefootnote}

\journal{Acta Astronautica}

\begin{document}
\begin{frontmatter}



\title{Low-Thrust Under-Actuated Satellite Formation Guidance and Control Strategies}


\author[1]{Ahmed Mahfouz}
\author[2]{Gabriella Gaias}
\author[3]{Florio {Dalla Vedova}}
\author[1,4]{Holger Voos}

\affiliation[1]{organization={SnT, University of Luxembourg},
            addressline={29, Avenue J.F Kennedy}, 
            city={Luxembourg},
            postcode={1855}, 
            country={Luxembourg}}

\affiliation[2]{organization={Department of Aerospace Science and Technology, Politecnico di Milano},
            addressline={34, via La Masa}, 
            city={Milan},
            postcode={20156}, 
            country={Italy}}

\affiliation[3]{organization={LuxSpace},
            addressline={9, Rue Pierre Werner}, 
            city={Betzdorf},
            postcode={6832}, 
            country={Luxembourg}}

\affiliation[4]{organization={Faculty of Science, Technology and Medicine, University of Luxembourg},
            addressline={2, place de l’Université}, 
            city={Esch-sur-Alzette},
            postcode={4365}, 
            country={Luxembourg}}

\begin{abstract}
This study presents autonomous guidance and control strategies for the purpose of reconfiguring close-range multi-satellite formations. The formation under consideration includes $N$ under-actuated deputy satellites and an uncontrolled virtual or physical chief spacecraft. Each deputy satellite is under-actuated due to the fact that it is equipped with a single low-thrust nozzle that is throttleable but ungimbaled. This setup requires deputies to perform maneuvers using a combination of thrust and coast arcs, during which attitude adjustments redirect the nozzle.
The guidance problem is formulated as a trajectory optimization problem that incorporates typical dynamical and physical constraints, alongside a minimum acceleration threshold. This latter constraint arises from the physical limitations of the adopted low-thrust technology, which is commonly employed for precise, close-range relative orbital maneuvers.
The guidance and control problem is addressed in two frameworks: centralized and distributed. The centralized approach provides a fuel-optimal solution, but it is practical only for formations with a small number of deputies. Conversely, the distributed approach is more scalable but yields sub-optimal solutions. In the centralized framework, the chief is a physical satellite responsible for all calculations, while in the distributed framework, the chief is treated as a virtual point mass orbiting the Earth, and each deputy performs its own guidance and control calculations onboard.
The study emphasizes the spaceborne implementation of the closed-loop control system, aiming for a reliable and automated solution to the optimal control problem. To this end, the risk of infeasibility is mitigated through first identifying the constraints that pose a potential threat of infeasibility, then properly softening them. Two Model Predictive Control architectures are implemented and compared, namely, a shrinking-horizon and a fixed-horizon schemes. Performances, in terms of fuel expenditure and achieved control accuracy, are analyzed on typical close-range reconfigurations requested by Earth observation missions and are compared against different implementations proposed in the literature.
\end{abstract}




\end{frontmatter}



\section{Introduction}\label{sec:Introduction}
Formation flying, where multiple satellites are coordinated to operate together on a shared mission, has gained significant importance in modern space missions due to its benefits in improving data quality, enhancing redundancy, and increasing mission flexibility \cite{Pelton2020SmallSatellites}. By deploying multiple small satellites in coordinated formations, these missions can achieve wider area coverage and provide more frequent data updates. Additionally, formation flying enables new possibilities in diverse applications, such as Earth observation, where multi-static Synthetic Aperture Radar (SAR) supports interferometry and tomography \cite{Krieger2010SAR_interferometry_FF, Fasano2014Formation_Geometry}.\\

The ability to autonomously and precisely control the positioning of satellites within a formation is a key technology for modern formation flying missions, improving operational flexibility and fulfilling mission requirements. This capability is particularly important for Earth observation missions, which require high precision over long mission durations. To achieve these goals efficiently, satellite formations have increasingly relied on electric propulsion technologies, which offer high control precision over extended periods while being generally more efficient than chemical thrusters \cite{Miller2021Survey}. Due to cost constraints, these missions often use small satellite platforms that require reductions in weight, volume, and power consumption of the satellite subsystems, including those of the propulsion subsystem. As a result, there is a growing trend to equip satellites with a single low-thrust nozzle, which provides a compact and efficient solution while still enabling the necessary control capabilities.\\

In this article, the problem of reconfiguring a formation of $N$ under-actuated satellites is considered, where the $N$ controlled satellites, referred to as deputy satellites, operate alongside an uncontrolled chief, which can be a physical or a virtual satellite. Formation reconfiguration in our context refers to the process of repositioning the deputies with respect to chief satellite given the initial and final configurations as well as the desired initial and final times of the maneuver.
The primary objective is to develop centralized and distributed guidance and control schemes that are:
\begin{enumerate}
    \item able to change the initial configuration into the final required one;
    \item Delta-V-optimal;
    \item compliant with operational constraints, such as maximum and minimum thrust limits as well as collision avoidance;
    \item suitable for onboard implementation in terms of computational efficiency;
    \item capable of autonomous operation either onboard the formation's central processing unit (the chief satellite) in the centralized setting or onboard each deputy in the distributed setting.
\end{enumerate}
The autonomy requirement aligns with the goals of the AuFoSat project, under which this research is conducted. AuFoSat aims to develop an autonomous Guidance, Navigation, and Control (GNC) library to support formation missions that comprise multiple Triton-X satellites. Previous AuFoSat research has covered topics including orbit design \cite{menzio2022formation}, relative navigation \cite{mahfouz2022relative, Mahfouz2023GNSS-based}, absolute orbit maintenance \cite{Mahfouz2023Autonomous}, relative orbit correction for two-satellite formations \cite{Mahfouz2024Fuel-Optimal}, and centralized guidance schemes for $N$-satellite formation reconfiguration \cite{mahfouz2024Delta_V_Optimal}. The latter article forms the foundation for the research presented here.\\

The choice to develop centralized schemes stems from the need to address limitations associated with distributed approaches, such as Delta-V sub-optimality and the potential for mis-coordination between satellites during collision-safe maneuvers \cite{Morgan14MPC}. Centralized strategies are particularly well-suited for small formations operating in close-range, as they deliver optimal solutions while maintaining simplicity in implementation. In contrast, distributed approaches excel in scalability, making them the preferred choice for formation missions involving a large number of deputy satellites.\\

The formation reconfiguration problem has been a point of significant research interest, particularly in the development of guidance and control methods. While numerous strategies have been tailored for formations using impulsive thrusters \cite{Larsson2011Autonomous, Gaias2015Impulsive, diMauro2018continuous, chernick2018new}, they do not translate effectively to formations utilizing low-thrust propulsion. Current low-thrust centralized frameworks often assume omnidirectional thrust capabilities \cite{scala2021design, DeVittori2022Low-Thrust}, a limitation that also applies to the distributed guidance and control strategies proposed in \cite{Morgan14MPC, Sarno2020PathPlanning}. As a result, these approaches are unsuitable for formations where each satellite is equipped with a single fixed thruster.\\

While the AVANTI mission developed guidance and control strategies for formations including a single impulsive-thrust satellite maneuvering around a passive target \cite{Gaias2018Avanti, Gaias2015Impulsive, Gaias2015Generalized}, these approaches are also not directly applicable to low-thrust applications or to $N$-satellite formations.
A Delta-V-optimal MPC strategy was introduced in \cite{Mahfouz2024Fuel-Optimal} for formations consisting of a single deputy and a chief satellite, specifically tailored for spacecraft equipped with a single low-thrust nozzle. However, this strategy did not incorporate minimum thrust limits or collision avoidance constraints, which are essential for operational safety and compliance. 
An alternative MPC approach was presented in \cite{Belloni2023Relative}, which tackled the problem of $N$-satellite formation reconfiguration with under-actuated deputies, incorporating both minimum thrust and collision avoidance constraints. Despite these additions, this scheme prioritized time efficiency over Delta-V optimality, potentially increasing fuel consumption. The same problem was further addressed in \cite{mahfouz2024Delta_V_Optimal}, where different efficient formulations for the guidance layer were developed, yet a fully closed-loop implementation was not provided.\\

Similar to many referenced works, the guidance task here is formulated as a trajectory optimization problem incorporating typical operational constraints such as collision avoidance, thrust limits, and dynamics constraints, while also allocating sufficient time for attitude maneuvers necessary to redirect the thruster nozzle. This research builds upon the work in \cite{mahfouz2024Delta_V_Optimal} by enhancing the guidance strategy used there with the addition of a minimum thrust constraint. The inclusion of this minimum thrust constraint poses a unique challenge due to its inherently non-convex nature. An affine convex approximation is therefore developed here, which calls for an initial estimate of thrust directions over the maneuver duration. This initial estimate is effectively achieved through a thrust pruning algorithm, which generates a reliable preliminary guess for thrust directions. A final enhancement to the guidance strategy is introduced, allowing the computational effort required to solve the guidance problem to be distributed among all deputy satellites. This distributed framework necessitates proper scheduling of the guidance calculations among the deputies to ensure coordinated motion.
Since the guidance scheme from \cite{mahfouz2024Delta_V_Optimal} serves as the foundation of the guidance layer employed here, the proposed approach in this work inherits several beneficial features of the reference guidance. Among these are the capacity to integrate extended no-thrust intervals within the reconfiguration maneuver, specifically during payload operations or when the deputies pass through eclipse periods. \\

To ensure robust closed-loop control, this research also employs two different MPC schemes that integrate the developed guidance. It further addresses the challenges of maintaining feasibility within the closed-loop system. Notably, the risk of encountering an infeasible optimization problem during MPC operation is mitigated by implementing a softening mechanism in the guidance layer, designed to handle infeasibility risks that may arise during optimization. This is achieved by relaxing critical constraints, namely those on the final state, minimum thrust, and collision avoidance. By introducing slack variables that permit slight, controlled violations of these constraints when necessary, the guidance layer ensures the MPC can consistently yield a feasible solution, preserving the maneuver’s integrity and feasibility.\\

The novelties introduced by this article include:
\begin{itemize}
    \item The development of a full closed-loop control system for the reconfiguration of formations comprising $N$ under-actuated satellites, with a focus on Delta-V optimality;
    \item Inclusion of the minimum thrust constraint through an affine convex approximation, supported by necessary algorithms such as the thrust pruning algorithm;
    \item Softening of critical constraints within the guidance layer to address potential infeasibility during MPC operations;
    \item Introduction of a scheduling logic to coordinate guidance calculations among deputies in the distributed setting, ensuring proper coordination between satellites.
\end{itemize}

This paper is structured as follows: Section \ref{sec:Dynamics} outlines the dynamics model governing the satellite formation. Section \ref{sec:Guidance} introduces the guidance strategies for reconfiguration, detailing the softening mechanism, thrust pruning technique, and methods to distribute the the computational efforts among deputies. In Section \ref{sec:Control}, the control schemes are discussed, with a focus on adaptations for onboard implementation. Section \ref{sec:Results} presents a comparative analysis of the two MPC schemes employed in both centralized and distributed settings, highlighting their respective strengths and trade-offs through simulation results across various reconfiguration scenarios. Finally, Section \ref{sec:Conclusion} provides concluding insights into the performance of each MPC scheme.

\section{Dynamics}\label{sec:Dynamics}
In this section the formulations used in describing the absolute as well as the relative orbital dynamics are briefly described, but firstly, the following reference frames are introduced:
\begin{itemize}
    \item The Earth-Centered Inertial (ECI) frame, which in our context is defined as the J2000 frame. A vector expressed in this frame is denoted by the superscript $\parenth{\cdot}^{i}$;
    \item The Radial-Transversal-Normal (RTN) frame, which is centered at the mass center of the chief. The x-axis of this frame is directed along the radial direction of the chief satellite and pointing away from the Earth, while its z-axis is normal to the orbital plane of the chief and the y-axis completes the right-handed triad. The superscript  $\parenth{\cdot}^{r}$ signifies that the vector is expressed in RTN frame.
\end{itemize}

The absolute state of a satellite can be described by the Cartesian state vector, $\vec{x}^{\parenth{\cdot}}$, collating the  position and velocity vectors of the spacecraft such that,
\begin{equation}
    \vec{x}^{\parenth{\cdot}} = \begin{bmatrix}
        \vec{r}^{\parenth{\cdot}} \\ \vec{v}^{\parenth{\cdot}}
    \end{bmatrix},
\end{equation}
where $\vec{r}^{\parenth{\cdot}}$ is position vector of a satellite described in the proper $\parenth{\cdot}$ frame and $\vec{v}^{\parenth{\cdot}}$ is its velocity vector.\\

An alternative way of describing the absolute state of a satellite is through the vector of quasi-non-singular orbital elements, defined as,
\begin{equation}
    \vec{\alpha} = \begin{bmatrix}
        a & \theta & e_{x} & e_{y} & i & \Omega
    \end{bmatrix}^{\intercal},
\end{equation}
where $a$ is the semi-major axis, $\theta$ is the mean argument of latitude, and $e_{x} = e \cos{\parenth{\omega}}$ and $e_{y} = e \sin{\parenth{\omega}}$ are the components of the eccentricity vector with $e$ being the orbital eccentricity and $\omega$ being the argument of perigee. Moreover, $i$ is the orbital inclination and $\Omega$ is the Right Ascension of the Ascending Node (RAAN). The vector $\vec{\alpha}$ includes the \emph{mean} quasi-non-singular orbital elements, where "mean" in our context refers to the one-orbit averaged states, ignoring the short- and long-term periodic effect of the second zonal harmonic of the Earth ($J_{2}$). Osculating orbital elements, which include the $J_{2}$-induced short- and long-term oscillations, are distinguished from the mean ones by an over tilde, i.e., $\tilde{\vec{\alpha}}$ represents the osculating orbital elements vector. The mapping between mean and osculating orbital elements, as well as the inverse transformation from osculating to mean elements, are outlined in \cite{Gaias2020Analytical}.\\

Similar to the absolute orbital motion, the relative state is parameterized by either the relative Cartesian state vector, $\Delta \vec{x}^{\parenth{\cdot}}$, or the quasi-non-singular Relative Orbital Elements (ROE), $\delta \vec{\alpha}$. The relative Cartesian states are simply the arithmetic difference between the Cartesian state vector of a deputy and that of the chief.
On the other hand, the quasi-non-singular ROE, or simply ROE in the rest of this text, are defined between a deputy and the chief as,
\begin{equation}\label{eq:ROE_def}
    \delta \vec{\alpha} = \begin{bmatrix} 
    \delta a \\
    \delta \lambda \\
    \delta e_{x} \\
    \delta e_{y} \\
    \delta i_{x} \\
    \delta i_{y} 
    \end{bmatrix} = \begin{bmatrix}
    \dfrac{\Delta a}{a_{c}} \\
    \Delta \theta + \Delta \Omega \cos{i_{c}} \\
    \Delta e_{x}\\
    \Delta e_{y} \\
    \Delta i \\
    \Delta \Omega \sin{i_{c}} 
    \end{bmatrix},
\end{equation}
where $\delta a$ is the relative semi-major axis, $\delta \lambda$ is the relative mean longitude, $\delta\vec{e}\coloneqq\begin{bmatrix} \delta e_{x} & \delta e_{y}\end{bmatrix}^{\intercal}$ is the relative eccentricity vector, and $\delta\vec{i}\coloneqq\begin{bmatrix} \delta i_{x} & \delta i_{y}\end{bmatrix}^{\intercal}$ is the relative inclination vector. Moreover, $a_{c}$ and $i_{c}$ are the semi-major axis and the inclination of the chief's orbit. 
It is to be noted that $\Delta \parenth{\cdot}$ signifies the arithmetic difference between a variable that relates to a deputy and that of the chief, while $\delta \parenth{\cdot}$ underpins a relative quantity between a deputy and the chief which is not necessarily the arithmetic difference between that of the deputy and that of the chief. Here, and in the rest of the text, a subscript $\parenth{\cdot}_{c}$ denotes a chief-related quantity, while a subscript $\parenth{\cdot}_{i}$ will refer to a variable that relates to the $i^\text{th}$ deputy.\\

The mean ROE vector, defined in \cref{eq:ROE_def}, is a dimensionless state vector. 
A dimensional ROE vector is obtained by multiplying the dimensionless ROE vector by the semi-major axis of the chief,
\begin{equation} \label{eq:y_def}
    \vec{y}_{i} = a_{c} \delta \vec{\alpha}_{i},
\end{equation}
where $\vec{y}_{i}$ is the dimensional mean ROE vector of the $i^\text{th}$ deputy, with units of length.\\

A linearized model for the dimensional ROE vector is described in \cite{diMauro2018continuous}. This model is written in a discrete form as,
\begin{equation}\label{eq:ROE_dynamics_sol}
    \vec{y}_{i} \parenth{t_{k+1}} = \mat{\Phi}\parenth{t_{k}, t_{k+1}} \vec{y}_{i} \parenth{t_{k}} + \mat{\Psi}\parenth{t_{k}, t_{k+1}} \overline{\vec{u}}_{i}^{r}\parenth{t_{k}, t_{k+1}}, 
\end{equation}
where $\mat{\Phi}\parenth{t_{k}, t_{k+1}}$ is the State Transition Matrix (STM) between the two time instances $t_{k}$ and $t_{k+1}$, $\mat{\Psi}\parenth{t_{k}, t_{k+1}}$ is the convolution matrix between the same two instances, and $\overline{\vec{u}}_{i}^{r}\parenth{t_{k}, t_{k+1}} = a_{c} {\vec{u}}_{i}^{r}\parenth{t_{k}, t_{k+1}}$ is the scaled control acceleration vector, with ${\vec{u}}_{i}^{r}\parenth{t_{k}, t_{k+1}}$ being the control acceleration vector; constant over the period $\left[t_{k}, t_{k+1}\right[$. To simplify the equations in the rest of the paper, the following notation is used; $\vec{y}_{i, k} \equiv \vec{y}_{i} \parenth{t_{k}}$, $\mat{\Phi}_{k} \equiv \mat{\Phi}\parenth{t_{k}, t_{k+1}}$, $\mat{\Psi}_{k} \equiv \mat{\Psi}\parenth{t_{k}, t_{k+1}}$, and $\overline{\vec{u}}_{i, k} \equiv \overline{\vec{u}}_{i}^{r}\parenth{t_{k}, t_{k+1}}$.\\

The guidance schemes developed in this work rely on propagating the relative dynamics in the ROE space, which does not grant a direct access to the Cartesian state vector at any given epoch. There exist, however, several non-linear mappings to relate the ROE to the relative Cartesian state vector \cite{Gaias2021Precise, Schaub2018FF}. For close-range near-circular formation flying, there exists a rigorous time-dependent transformations between a ROE vector and its corresponding relative Cartesian state in the RTN frame \cite{gaias2021trajectory}. This latter transformation is adopted in this paper and it is written as,
\begin{equation}
    \Delta \vec{r}^{r}_{i, k} = \mat{T}_{k} \vec{y}_{i,k},
\end{equation}
where $\mat{T}_{k} \equiv \mat{T} \parenth{t_{k}} \in \set{R}^{3 \times 6}$ is a time-dependent mapping matrix, which is described in details in \cite{gaias2021trajectory}.

\section{Guidance}\label{sec:Guidance}
This section begins by presenting a reference centralized guidance scheme from the literature, designed to optimize the relative trajectories of $N$ under-actuated deputy satellites flying in formation with an uncontrolled chief. The functionality of this guidance scheme is subsequently extended by incorporating a minimum acceleration threshold constraint. To further enhance the autonomy and robustness of the overall closed-loop control system, strategies are discussed to address the risk of infeasibility within the guidance layer of the proposed optimal control problem. Additionally, methods for distributing the computational effort of generating guidance profiles among all deputies are introduced to improve the scalability of the proposed guidance methods. This distributed approach contrasts with the centralized configuration, where guidance profiles are computed entirely on a central processing unit located on the uncontrolled chief.

\subsection{Reference guidance}
In \cite{mahfouz2024Delta_V_Optimal}, the guidance problem for an $N$-satellite formation reconfiguration was formulated as a trajectory optimization problem, where each of the $N$ satellites is equipped with a single electric thruster nozzle. In that article, $4$ different convex formulations have been proposed, which vary in terms of the transcription of the optimal control problem, Delta-V requirements, and required time to solve. It was concluded in \cite{mahfouz2024Delta_V_Optimal} that casting the guidance as a Second-Order Cone Programming (SOCP) problem is generally recommended over the other proposed formulations for being the 
most efficient from the point of view of fuel consumption as well as the required time to solve it by most of the $15$ solvers compared in \cite{mahfouz2024Delta_V_Optimal}.\\

\begin{figure*}[ht]
    \centering
    \includegraphics[width=\linewidth]{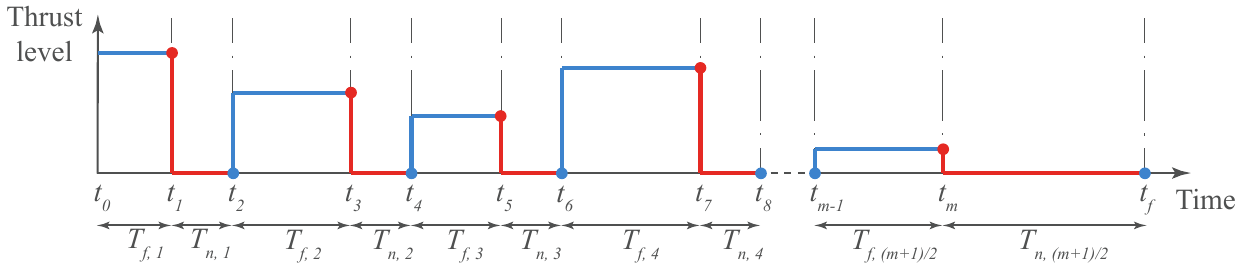}
    \caption{Graphical representation of the low-thrust guidance strategy}
    \label{fig:Low-thrust-guidance-scheme}
\end{figure*}

The SOCP formulation of the guidance problem, as formulated in \cite{mahfouz2024Delta_V_Optimal}, is presented in this paper for completeness, but before that, a brief preamble needs to be introduced. Firstly, the problem that we consider here is the change of the initial configuration of an under-actuated $N$-satellite formation at the initial time of the maneuver $t_{0}$ to a desired configuration at the user-defined final time of the maneuver $t_{f}$. In order to address the problem of the deputy satellites being each equipped with an ungimbaled single low-thrust nozzle, the maneuver time was divided into multiple control cycles, where each cycle comprises a thrust and a coast arc. An attitude maneuver is allocated during each of the coast arcs in order to redirect the thruster nozzle to the desired thrust direction for the respective subsequent thrust arc. The maneuver is performed through exactly $\dfrac{m+1}{2}$ control cycles, where $m$ is an odd number. The concept of the repeated control cycles is illustrated graphically in \cref{fig:Low-thrust-guidance-scheme}. 
Indeed, the control cycles need not to be of the same length. In fact, the time instances at which the thruster is turned on and off are user-defined.
The fact that these time instances are left for the user to determine is motivated by supporting the predictability of the maneuver, which many operators favor over the fuel or time optimality. Additionally, letting the user define these time instances allows the trajectory optimization strategy to be cast as a convex optimization problem, and also allows for accommodating periods where thrusters are not allowed to fire, e.g., during eclipse or during ground contact.

It is important to note that the indices for the forced motion time instances, representing the start of each thrust arc, are collated in the set $\dist{K}_{f} = \curlyb{0, 2, 4, \hdots, m-1}$, while the indices of the natural motion time instances, relating to the beginning of a coast arc, are put together in the set $\dist{K}_{n} = \left\{1, 3, 5 \hdots, m\right\}$.\\

Given the aforementioned architecture according to which the trajectory optimization problem is handled, the SOCP guidance problem proposed by \cite{mahfouz2024Delta_V_Optimal} can then be written as,
\begin{Problem}[SOCP formulation]
\label{prob:SOCP_formulation}
\begin{align}
& \min_{\mat{Y}, \overline{\mat{U}}, \mat{\Gamma}} \quad \frac{1}{a_{c}}\sum_{i\in \dist{I}}\sum_{k \in \dist{K}_{f}}{\parenth{\Delta t_{k} \Gamma_{i, k}}} \nonumber\\
& \text{subject to,} \nonumber\\
& \vec{y}_{i, 0} = \vec{y}_{i, 0}, \qquad \vec{y}_{i, m+1} = \overline{\vec{y}}_{i, f} \quad \forall i \in \dist{I}, \label{eq:boundary_constraints_SOCP}\\
& \vec{y}_{i, k+1} =  \mat{\Phi}_{k} \vec{y}_{i, k} + \mat{\Psi}_{k} \overline{\vec{u}}_{i, k} \quad \forall i \in \dist{I} ,\; \forall k \in \dist{K},\label{eq:dynamics_constraint_SOCP}\\
& \overline{\vec{u}}_{i, k} = \vec{0} \quad \forall i \in \dist{I} ,\; \forall k \in \dist{K}_{n}, \label{eq:u0_constraint_SOCP}\\
& \norm{\overline{\vec{u}}_{i, k}} \leq \Gamma_{i,k}, \quad \forall i \in \dist{I} ,\; \forall k \in \dist{K}_{f}, \label{eq:umax_constraint_SOCP}\\
& \Gamma_{i,k} \leq a_{c} u_{i, \text{max}} \quad \forall i \in \dist{I},\; \forall k \in \dist{K}_{f},\\
& \begin{multlined}[t][\multlinedwidth]
    \dfrac{\parenth{\breve{\vec{y}}_{i, k} - \breve{\vec{y}}_{j, k}}^{\intercal}}{\norm{\mat{T}_{k}\parenth{\breve{\vec{y}}_{i, k} - \breve{\vec{y}}_{j, k}}}}
    \mat{T}_{k}^{\intercal} \mat{T}_{k} \parenth{\vec{y}_{i, k} - \vec{y}_{j, k}} \geq R_\text{CA} \\
    \hfill \forall i, j \in \dist{I} ,\; i \neq j ,\; \forall k \in \dist{K},
\end{multlined} \label{eq:CA_deputy_deputy_SOCP}\\
& \dfrac{\breve{\vec{y}}_{i, k}^{\intercal}}{\norm{\mat{T}_{k} \breve{\vec{y}}_{i, k}}} \mat{T}_{k}^{\intercal} \mat{T}_{k} \vec{y}_{i, k} \geq  R_\text{CA} \quad \forall i \in \dist{I} ,\; \forall k \in \dist{K},\label{eq:CA_deputy_chief_SOCP}
\end{align}
\end{Problem}
where $\mat{Y}$, $\overline{\mat{U}}$, and $\mat{\Gamma}$ are the matrices which collate the state profile, the control profile, and the slack variables $\Gamma_{i, k}$, respectively, $\dist{I} = \curlyb{1, 2, \hdots, N}$ is the set of deputies' indices, $\dist{K} = \dist{K}_{f} \cup \dist{K}_{n}$ is the set of all time indices except the final time instance, and $\Delta t_{k} = t_{k+1} - t_{k}$. Moreover, $\vec{y}_{i,0}$ is the dimensional ROE vector of the $i^\text{th}$ deputy at the initial time of the maneuver $t_{0}$,  $\overline{\vec{y}}_{i,f}$ is the user-defined desired state of the $i^\text{th}$ satellite at the final time of the maneuver $t_{f}$, $u_{i, \text{max}}$ is the maximum allowable acceleration by the onboard thruster of the $i^\text{th}$ deputy, and $R_{CA}$ is the radius of the Keep-Out Zone (KOZ) around each satellite, which in our context is assumed a sphere. It is important to emphasize that the vectors $\breve{\vec{y}}_{i,k}$ represent the initial guess for the state profile, and a method to obtain them will be discussed later. The matrices  $\mat{Y}$, $\overline{\mat{U}}$, and $\mat{\Gamma}$ are formally constructed as,
\begin{equation}
\begin{split}\label{eq:Y_U_Gamma_def}
    \mat{Y} & = \brack{\brack{\vec{y}_{i, k}}^{h}_{i\in \dist{I}}}^{v}_{k \in \dist{K} \cup \curlyb{m+1}},\\  &\\ 
    \overline{\mat{U}} & = \brack{\brack{\overline{\vec{u}}_{i, k}}^{h}_{i\in \dist{I}}}^{v}_{k \in \dist{K}},\\
    &\\
    \mat{\Gamma} & = \brack{\brack{\Gamma_{i, k}}^{h}_{i\in \dist{I}}}^{v}_{k \in \dist{K}_{f}},\\
\end{split}
\end{equation}
where $\brack{\parenth{\cdot}_{i}}^{h}_{\text{indices}}$ and $\brack{\parenth{\cdot}_{i}}^{v}_{\text{indices}}$ are defined as the horizontal and the vertical concatenation operators, respectively, such that $\brack{a_{i}}^{h}_{i\in \curlyb{1, 2, \hdots, n}} = \begin{bmatrix} a_{1} & a_{2} & \hdots & a_{n} \end{bmatrix}$ and $\brack{a_{i}}^{v}_{i\in \curlyb{1, 2, \hdots, n}} = \begin{bmatrix} a_{1} & a_{2} & \hdots & a_{n} \end{bmatrix}^{\intercal}$.\\

The cost function of Problem \ref{prob:SOCP_formulation} represents the total Delta-V required by all the deputy satellites to complete the formation reconfiguration maneuver, while the constraints imposed on the problem are summarized as follows,
\begin{itemize}
    \item \cref{eq:boundary_constraints_SOCP} enforces boundary conditions, ensuring the final state matches user-defined set points while respecting each deputy's initial state;
    \item The trajectory optimization adheres to the relative orbital dynamics via \cref{eq:dynamics_constraint_SOCP};
    \item Maximum acceleration limits are enforced through \cref{eq:umax_constraint_SOCP}, while \cref{eq:u0_constraint_SOCP} ensures zero acceleration during coast arcs;
    \item Collision avoidance is addressed through \cref{eq:CA_deputy_deputy_SOCP,eq:CA_deputy_chief_SOCP}, preventing inter-deputy collisions and deputy-chief collisions at each time step.
\end{itemize}

It is to be noted that Problem \ref{prob:SOCP_formulation} requires the knowledge of $\breve{\vec{y}}_{i, k} \; \forall i \in \dist{I},\; \forall k \in \dist{K}$. This is treated through the adoption of a Sequential Convex Programming (SCP) scheme, in which the problem is first solved without the collision avoidance constraints, \cref{eq:CA_deputy_deputy_SOCP,eq:CA_deputy_chief_SOCP}, and then the problem is solved multiple times until one of the termination criteria is satisfied. Indeed, $\breve{\vec{y}}_{i, k}$ for the current iteration is set to the solution $\vec{y}_{i, k}$ from the previous iteration. In this article, and in \cite{mahfouz2024Delta_V_Optimal}, three termination criteria are adopted. Namely:
\begin{itemize}
    \item $\norm{\breve{\vec{y}}_{i, k} - \vec{y}_{i, k}}\leq\epsilon$ at the current iteration, with $\epsilon>0$ being a user-defined threshold; 
    \item The guidance profile of the current iteration is collision-free; 
    \item The user-defined iteration limit is reached, in which case, the solution trajectory is not guaranteed to be collision-free.
\end{itemize}
One of the main advantages of the employed termination criteria for the SCP strategy is that the guidance problem typically needs to be solved only one time after the zeroth iteration, i.e., the iteration in which the problem is solved without the collision avoidance constraints. Although this results in a sub-optimal solution, it has been illustrated in \cite{mahfouz2024Delta_V_Optimal} that the obtained solution is very close to the optimal one. It is important to emphasize that obtaining the optimal solution generally requires several SCP iterations before it can be obtained.\\

\begin{figure}[ht]
    \centering
    \includegraphics[width=\columnwidth]{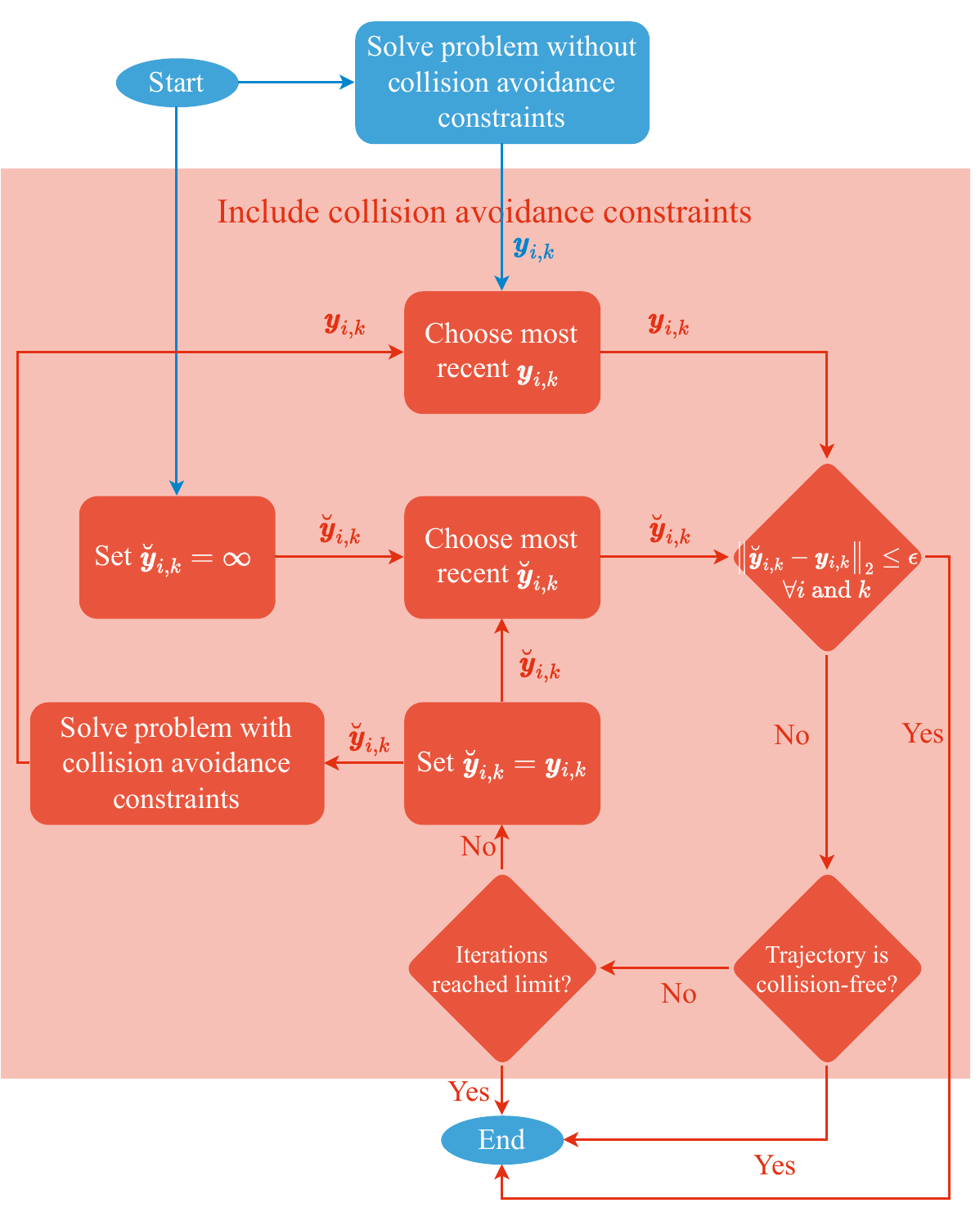}
    \caption{Graphical representation of the SCP scheme}
    \label{fig:SCP_scheme}
\end{figure}

A schematic illustrating the adopted flow of the SCP scheme is presented in \cref{fig:SCP_scheme}.

\subsection{Inclusion of the Minimum thrust constraint}
The minimum thrust/acceleration threshold represents an additional constraint not considered in Problem \ref{prob:SOCP_formulation}, yet it significantly impacts the accuracy of the relative control system. From a practical perspective, maneuvers that require thrust levels below a certain threshold cannot be executed by the Hall Effect thruster used in Triton-X, nor by many Commercial Off-The-Shelf (COTS) electric thrusters. To this end, the aspect of the lower thrust/acceleration bound is directly addressed in the guidance problem formulation to ensure that the reconfiguration is optimized for practical feasibility and can be realistically achieved. In the context of our low-thrust formation reconfiguration problem, where each deputy is equipped with a single thruster, this bound translates to a constraint on the $L_{2}$ norm of the control thrust vector, or alternatively, on the $L_{2}$ norm of the control acceleration vector, assuming a constant mass of the satellite throughout the maneuver. This can be formally written as,
\begin{equation}\label{eq:min_acc_constraint}
    \norm{\vec{u}^{r}_{i, k}} \geq u_{i, \text{min}},
\end{equation}
where $u_{i, \text{min}}$ the minimum allowable acceleration level by the $i^\text{th}$ deputy. \\

\begin{figure*}[ht]
\begin{subfigure}[c]{\columnwidth}
    \centering
    \includegraphics[width=0.7\columnwidth]{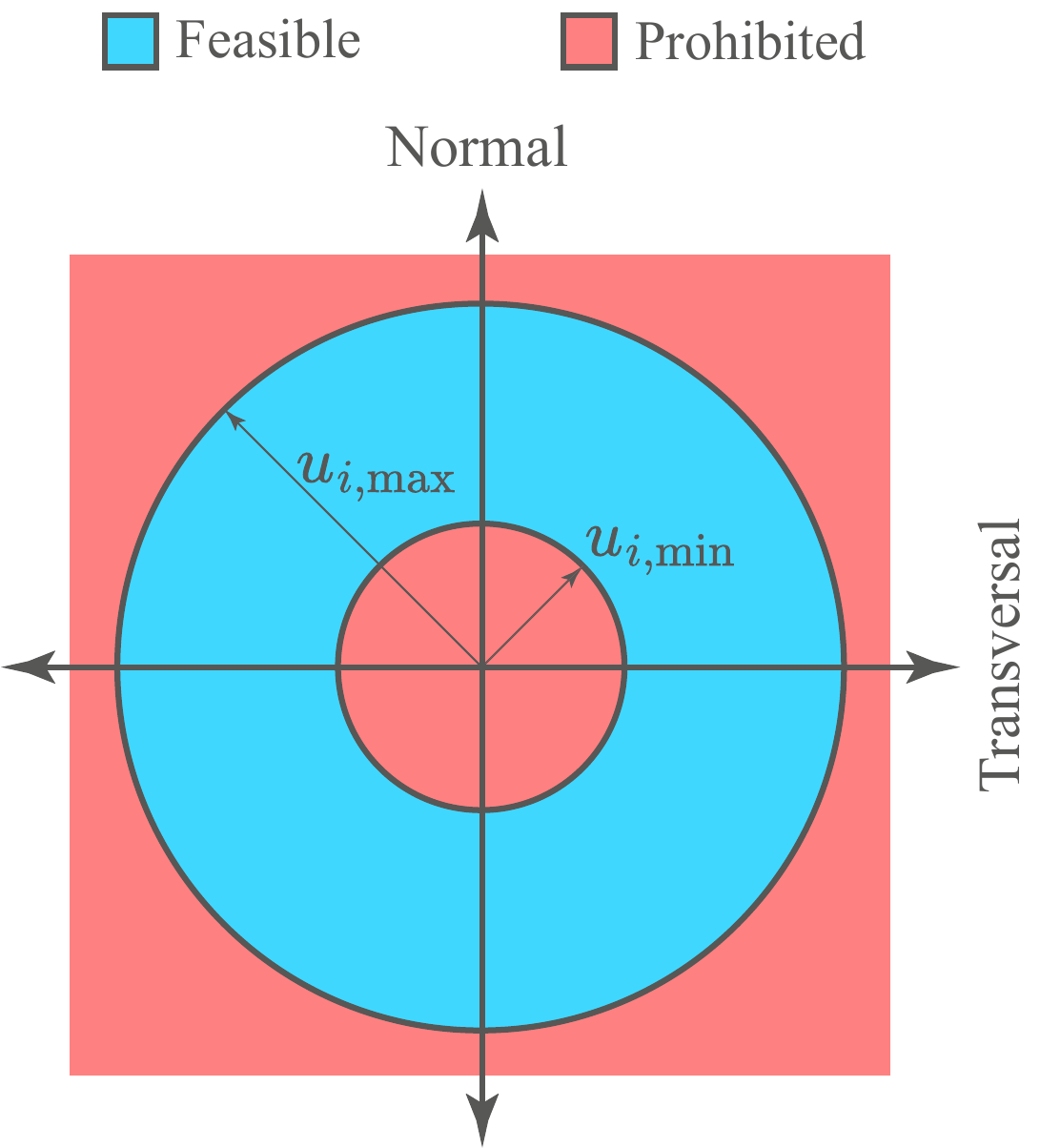}
    \caption{Original non-convex constraint} 
\end{subfigure}
\hfill
\begin{subfigure}[c]{\columnwidth}
    \centering
    \includegraphics[width=0.7\columnwidth]{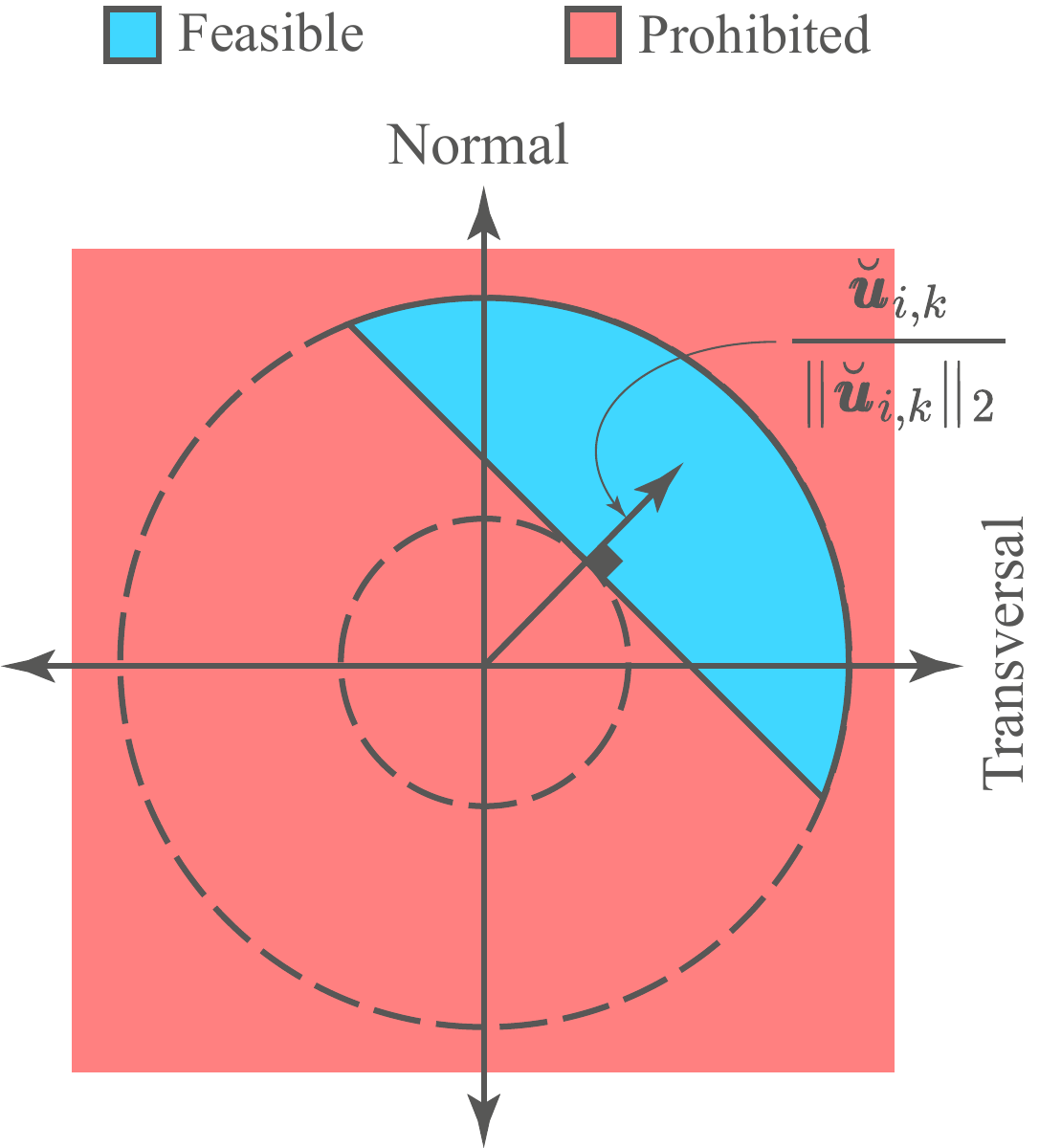}
    \caption{Relaxed convex constraint}
\end{subfigure}
\caption{Feasibility region of the acceleration vector}
\label{fig:control_feasibility_region_with_umin}
\end{figure*}

From a mathematical standpoint, the constraint in \cref{eq:min_acc_constraint} is a non-convex one, since it represents the outer region of a sphere with radius $u_{i, \text{min}}$. If the minimum acceleration constraint is incorporated in Problem \ref{prob:SOCP_formulation}, it directly renders it non-convex.
A lossless convexification technique was proposed \cite{Acikmese2011Lossless, Blackmore2012Lossless} for this non-convex constraint, yet the biggest drawback of this convexification proposal is that it relies on specific observability conditions, which are not met by the ROE dynamics. Namely, when the ROE dynamics are expressed in the standard state-space form as a Linear-Time-Varying (LTV) system, the pair comprising the system matrix and the input matrix must be fully observable, or equivalently, the system must be fully controllable. However, as demonstrated in \cite{Sasaki2024ROEControllability}, this is not true for the ROE dynamics.
Here, the minimum acceleration constraint is convexified using a dynamics-independent affine relaxation, where the constraint in \cref{eq:min_acc_constraint} is rewritten in an approximated convex form as follows:
\begin{equation}\label{eq:min_acc_constraint_relaxed}
     \breve{\vec{u}}_{i, k}^{\intercal} \vec{u}^{r}_{i, k} \geq \norm{\breve{\vec{u}}_{i, k}} u_{i, \text{min}},
\end{equation}
where $\breve{\vec{u}}_{i, k}$ is assumed to be a known vector. A visual representation of this relaxation, using the projections of $\breve{\vec{u}}_{i, k}$ and $\vec{u}^{r}_{i, k}$ into the T-N plane without loss of generality, is depicted \cref{fig:control_feasibility_region_with_umin} together with the original non-convex constraint. The maximum acceleration constraint is also included in the figure for completeness.

The relaxation in \cref{eq:min_acc_constraint_relaxed} can be rewritten in terms of $\overline{\vec{u}}_{i, k} = a_{c} \vec{u}^{r}_{i, k}$, so that it could be added to the list of constraints of Problem \ref{prob:SOCP_formulation} as,
\begin{equation}\label{eq:min_acc_bar_constraint_relaxed}
     \breve{\overline{\vec{u}}}_{i, k}^{\intercal} \overline{\vec{u}}_{i, k} \geq \norm{\breve{\overline{\vec{u}}}_{i, k}} a_{c} u_{i, \text{min}},
\end{equation}
where, again, $\breve{\overline{\vec{u}}}_{i, k}$ is assumed to be a known vector.\\

\begin{figure*}[ht]
    \centering
    \includegraphics[width=0.7\linewidth]{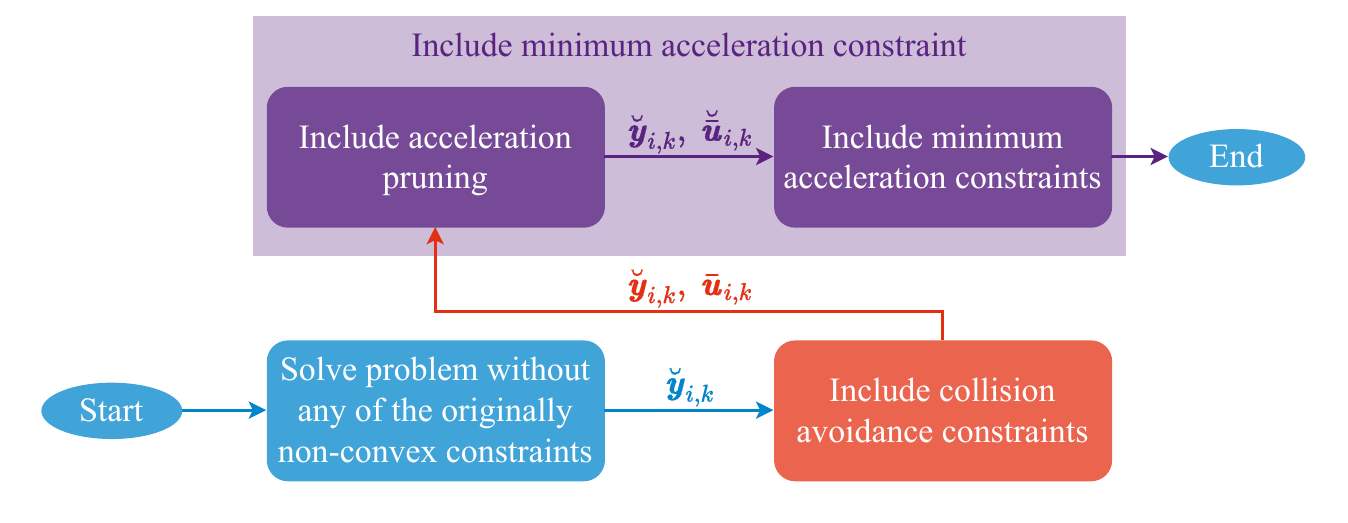}
    \caption{Graphical representation of the SCP scheme with minimum acceleration constraint}
    \label{fig:SCP_scheme_umin}
\end{figure*}

One of the most significant challenges in implementing the proposed relaxation in \eqref{eq:min_acc_bar_constraint_relaxed} lies in determining the linearization direction, denoted as ${\breve{\overline{\vec{u}}}_{i,k}}/{\norm{\breve{\overline{\vec{u}}}_{i,k}}}$. This task is particularly complex because the linearization direction must be computed individually for each satellite and at every discrete time step, effectively acting as an initial guess for the affine relaxation process. The selection of this direction is critical, as it directly influences the feasibility and efficiency of the convexified optimization problem.
To generate this initial guess, a two-step approach is proposed. The first step involves solving the problem without enforcing the minimum acceleration constraint given in \eqref{eq:min_acc_bar_constraint_relaxed}. By doing so, an unconstrained solution for the acceleration profile, $\vec{u}_{i,k}$, is obtained for each satellite. This step namely incorporates the SCP scheme in \cref{fig:SCP_scheme}. 
In the second step, an \emph{acceleration pruning process} is applied. This process begins by analyzing the acceleration profile of each satellite. Specifically, if the mean acceleration over the entire profile is found to be less than a pre-specified threshold, typically $1.5 \, u_{i, \text{min}}$, then the instances corresponding to the lowest accelerations are constrained to zero. This step is justified by the fact that the applied acceleration can, in practice, take values anywhere between the minimum and maximum allowable levels, and can also be zero. By selectively setting the smallest accelerations to zero, and re-solving the problem, the remaining time instances can utilize greater levels of acceleration. This redistribution better satisfies the minimum acceleration constraint and leads to a solution that is likely to be feasible when \cref{eq:min_acc_bar_constraint_relaxed} is imposed at a later stage.
Once the pruning process is complete, the pruned acceleration profile serves as the initial guess for the linearization direction in the affine relaxation. This refined guess is then employed in subsequent iterations of the optimization problem, which now includes the convexified version of the minimum acceleration constraint, \cref{eq:min_acc_bar_constraint_relaxed}. By incorporating these steps into the SCP strategy, \cref{fig:SCP_scheme} is modified, and the new version is depicted in \cref{fig:SCP_scheme_umin}.

It is important to note that, when the minimum acceleration constraints are included, the collision avoidance constraints are also taken into account. The results from the previous iteration are used as an initial guess for the affine relaxation of the collision avoidance constraints.\\

The pruning process is formally introduced in Algorithm \ref{alg:acceleration_pruning}, where the sets of indices $\dist{K}_{i}$ are identified for each deputy. 
These indices correspond to the time steps with the lowest acceleration values. The cardinality of each $\dist{K}_{i}$ is determined based on the user-defined pruning factors, $p_{i}$, as well as the mean value of the acceleration profile for the deputy satellite in question, as illustrated by Algorithm \ref{alg:acceleration_pruning}.

\begin{algorithm}
\DontPrintSemicolon
\SetAlgoLined
Solve the problem without the minimum acceleration constraint; Obtain the solution $\vec{u}_{i, k}$\;
  \ForEach{$i\in \dist{I}$}{
      $\dist{U}_{i} \gets \curlyb{\norm{\vec{u}_{i, 0}}, \norm{\vec{u}_{i, 2}}, \hdots, \norm{\vec{u}_{i, m-1}}}$\;
      $u_{i, \text{mean}} \gets \text{mean}\parenth{\dist{U}_{i}}$\;
      \uIf{$\parenth{u_{i, \textnormal{mean}} \leq 1.5\; u_{i, \textnormal{min}}}$}{
      $n_{i} \gets \lfloor p_{i}\cdot \parenth{1- \frac{u_{i, \text{mean}}}{u_{i, \text{min}}} \cdot \abs{\dist{K}_{f}}}\rfloor$\;
      \uIf{$\parenth{n_{i} > \abs{\dist{K}_{f}} - 2}$}{
        $n_{i} \gets \abs{\dist{K}_{f}} - 2$\;
      }
      $\dist{K}_{i} \gets$ Set of $\parenth{k}$ indices corresponding to the lowest $n_{i}$ values in $\dist{U}_{i}$\;
      }
  }
\caption{Acceleration pruning}
\label{alg:acceleration_pruning}
\end{algorithm}

Note that in Algorithm \ref{alg:acceleration_pruning}, $\abs{\dist{K}_{f}}$ represents the cardinality of the set $\dist{K}_{f}$, while $p_{i}$ is the pruning factor of the $i^\text{th}$ deputy. Moreover, the condition in line 7 is set to guarantee that at least two acceleration instances are not pruned. This is important because maintaining at least two control instances can be necessary since the action taken by the first acceleration instance, e.g., adjusting the value of $\delta a$ in order to achieve a desired $\delta \lambda$ behaviour, might need to be corrected at a later point in the profile.
Once the indices $\dist{K}_{i}$ are determined, they are used in solving the re-formulated SOCP scheme, which is presented in Problem \ref{prob:SOCP_formulation_pruning}.
\begin{Problem}[SOCP with acceleration pruning]
\label{prob:SOCP_formulation_pruning}
\begin{align}
& \min_{\mat{Y}, \overline{\mat{U}}, \mat{\Gamma}} \quad \frac{1}{a_{c}}\sum_{i\in \dist{I}}\sum_{k \in \dist{K}_{f} \setminus \dist{K}_{i}}{\parenth{\Delta t_{k} \Gamma_{i, k}}} \nonumber\\
& \text{subject to,} \nonumber\\
& \vec{y}_{i, 0} = \vec{y}_{i, 0}, \qquad \vec{y}_{i, m+1} = \overline{\vec{y}}_{i, f} \quad \forall i \in \dist{I}, \label{eq:boundary_constraints_SOCP_pruning}\\
& \vec{y}_{i, k+1} =  \mat{\Phi}_{k} \vec{y}_{i, k} + \mat{\Psi}_{k} \overline{\vec{u}}_{i, k} \quad \forall i \in \dist{I} ,\; \forall k \in \dist{K},\label{eq:dynamics_constraint_SOCP_pruning}\\
& \overline{\vec{u}}_{i, k} = \vec{0} \quad \forall i \in \dist{I},\; \forall \parenth{i, k} \in \parenth{i, \dist{K}_{n} \cup \dist{K}_{i}},  \label{eq:u0_constraint_SOCP_pruning}\\
& \norm{\overline{\vec{u}}_{i, k}} \leq \Gamma_{i,k} \quad \forall i \in \dist{I} ,\; \forall \parenth{i, k} \in \parenth{i, \dist{K}_{f} \setminus \dist{K}_{i}}, \label{eq:umax_constraint_SOCP_pruning}\\
& \Gamma_{i,k} \leq a_{c} u_{i, \text{max}} \quad \forall i \in \dist{I},\; \forall \parenth{i, k} \in \parenth{i, \dist{K}_{f} \setminus \dist{K}_{i}},\\
& \begin{multlined}[t][\multlinedwidth]
    \dfrac{\parenth{\breve{\vec{y}}_{i, k} - \breve{\vec{y}}_{j, k}}^{\intercal}}{\norm{\mat{T}_{k}\parenth{\breve{\vec{y}}_{i, k} - \breve{\vec{y}}_{j, k}}}}
    \mat{T}_{k}^{\intercal} \mat{T}_{k} \parenth{\vec{y}_{i, k} - \vec{y}_{j, k}} \geq R_\text{CA} \\
    \hfill \forall i, j \in \dist{I} ,\; i \neq j ,\; \forall k \in \dist{K},
\end{multlined} \label{eq:CA_deputy_deputy_SOCP_pruning}\\
& \dfrac{\breve{\vec{y}}_{i, k}^{\intercal}}{\norm{\mat{T}_{k} \breve{\vec{y}}_{i, k}}} \mat{T}_{k}^{\intercal} \mat{T}_{k} \vec{y}_{i, k} \geq  R_\text{CA} \quad \forall i \in \dist{I} ,\; \forall k \in \dist{K},\label{eq:CA_deputy_chief_SOCP_pruning}
\end{align}
\end{Problem}

As indicated by \cref{fig:SCP_scheme_umin}, the solution of Problem \ref{prob:SOCP_formulation_pruning} is used as an initial guess for the minimum thrust constraint in \cref{eq:min_acc_bar_constraint_relaxed}. The full guidance problem, including the minimum thrust constraint, is written in the SOCP form in Problem \ref{prob:SOCP_formulation_umin}.
\begin{Problem}[SOCP with minimum acceleration cons.]
\label{prob:SOCP_formulation_umin}
\begin{align}
& \min_{\mat{Y}, \overline{\mat{U}}, \mat{\Gamma}} \quad \frac{1}{a_{c}}\sum_{i\in \dist{I}}\sum_{k \in \dist{K}_{f} \setminus \dist{K}_{i}}{\parenth{\Delta t_{k} \Gamma_{i, k}}} \nonumber\\
& \text{subject to,} \nonumber\\
& \vec{y}_{i, 0} = \vec{y}_{i, 0}, \qquad \vec{y}_{i, m+1} = \overline{\vec{y}}_{i, f} \quad \forall i \in \dist{I}, \label{eq:boundary_constraints_SOCP_umin}\\
& \vec{y}_{i, k+1} =  \mat{\Phi}_{k} \vec{y}_{i, k} + \mat{\Psi}_{k} \overline{\vec{u}}_{i, k} \quad \forall i \in \dist{I} ,\; \forall k \in \dist{K},\label{eq:dynamics_constraint_SOCP_umin}\\
& \overline{\vec{u}}_{i, k} = \vec{0} \quad \forall i \in \dist{I},\; \forall \parenth{i, k} \in \parenth{i, \dist{K}_{n} \cup \dist{K}_{i}},  \label{eq:u0_constraint_SOCP_umin}\\
& \begin{multlined}[t][\multlinedwidth]
    \breve{\overline{\vec{u}}}_{i, k}^{\intercal} \overline{\vec{u}}_{i, k} \geq \norm{\breve{\overline{\vec{u}}}_{i, k}} a_{c} u_{i, \text{min}} \\ \forall i \in \dist{I} ,\; \forall \parenth{i, k} \in \parenth{i, \dist{K}_{f} \setminus \dist{K}_{i}},
\end{multlined} \label{eq:umin_constraint_SOCP_umin}\\
& \norm{\overline{\vec{u}}_{i, k}} \leq \Gamma_{i,k} \quad \forall i \in \dist{I} ,\; \forall \parenth{i, k} \in \parenth{i, \dist{K}_{f} \setminus \dist{K}_{i}}, \label{eq:umax_constraint_SOCP_umin}\\
& \Gamma_{i,k} \leq a_{c} u_{i, \text{max}} \quad \forall i \in \dist{I},\; \forall \parenth{i, k} \in \parenth{i, \dist{K}_{f} \setminus \dist{K}_{i}},\\
& \begin{multlined}[t][\multlinedwidth]
    \dfrac{\parenth{\breve{\vec{y}}_{i, k} - \breve{\vec{y}}_{j, k}}^{\intercal}}{\norm{\mat{T}_{k}\parenth{\breve{\vec{y}}_{i, k} - \breve{\vec{y}}_{j, k}}}}
    \mat{T}_{k}^{\intercal} \mat{T}_{k} \parenth{\vec{y}_{i, k} - \vec{y}_{j, k}} \geq R_\text{CA} \\
    \hfill \forall i, j \in \dist{I} ,\; i \neq j ,\; \forall k \in \dist{K},
\end{multlined} \label{eq:CA_deputy_deputy_SOCP_umin}\\
& \dfrac{\breve{\vec{y}}_{i, k}^{\intercal}}{\norm{\mat{T}_{k} \breve{\vec{y}}_{i, k}}} \mat{T}_{k}^{\intercal} \mat{T}_{k} \vec{y}_{i, k} \geq  R_\text{CA} \quad \forall i \in \dist{I} ,\; \forall k \in \dist{K},\label{eq:CA_deputy_chief_SOCP_umin}
\end{align}
\end{Problem}

While solving Problem \ref{prob:SOCP_formulation_umin} may yield feasible solutions for some reconfiguration scenarios, there is no guarantee that it will always be feasible for other reconfiguration tasks. Unlike Problem \ref{prob:SOCP_formulation}, which is guaranteed to provide a feasible solution given a sufficiently large maneuver time, Problem \ref{prob:SOCP_formulation_umin} is heavily dependent on the choice of linearization directions $\breve{\overline{\vec{u}}}_{i, k}$. If these directions are poorly selected, the problem is prone to infeasibility, even when the allowed maneuver time is extended. Indeed, in view of the spaceborne implementation of the proposed guidance, it is crucial to minimize the likelihood of encountering unsolvable scenarios. The following discussion outlines techniques designed to mitigate the risk of infeasibility.

\subsection{Infeasibility handling}
To support autonomy of the formation control tasks, the proposed guidance strategy is meant to be solved recurrently within an MPC scheme, which makes use of the available sensor measurements to perform the control task in an efficient manner while accounting for the disturbances and model uncertainties. It does so through optimizing the trajectory as well as the control profile over a period of time (horizon), then applies the optimized control profile up to a certain point (usually only the first control step is utilized). The details of two distinct MPC strategies are discussed later in \cref{sec:Control}. 
One major risk for an MPC is arriving to a point where the guidance optimization problem, on which the MPC heavily depends, is infeasible. The mitigation of this risk is handled in our context inside the guidance layer, where all the constraints that may lead to infeasibility are softened in order to ensure that a feasible solution is always available for the MPC. This comes understandably at the cost of slight violations of the constraints if the problem is not initially feasible.\\

Before introducing methods to soften the constraints, those which are expected to lead to infeasibility must  be first identified.
As discussed earlier, the relaxed minimum thrust constraint, in \cref{eq:umin_constraint_SOCP_umin}, is responsible for potential infeasibility because of its dependence on the initial guess, $\breve{\overline{\vec{u}}}_{i, k}$. Indeed, a good guess for $\breve{\overline{\vec{u}}}_{i, k}$ is obtained through the acceleration pruning process, nonetheless, infeasibility could still be expected mainly because the optimizer is not able to find a solution that satisfies the final state constraint, in \cref{eq:boundary_constraints_SOCP_umin}, while respecting the dynamics constraints in \cref{eq:dynamics_constraint_SOCP_umin} and at the same time being constrained by the thrust directions imposed by the constraints in \cref{eq:umin_constraint_SOCP_umin}. It is for this reason that both, the final state constraints (\cref{eq:boundary_constraints_SOCP_umin}), and the relaxed minimum acceleration constraints (\cref{eq:umin_constraint_SOCP_umin}), are softened to reduce the risk of infeasibility. The process of softening the final state constraints involves broadening the original objective of these constraints. Initially, the goal was to align the final state of the optimized profile with the desired one. The revised objective, however, is for the state profile to approximate a user-defined profile at specific user-defined time instances. This approach is supported by the fact that many MPC schemes, not to mention the Fixed-Horizon MPC (FHMPC) that will be introduced in \cref{sec:Control}, rely on tracking a user-defined state profile which typically results from solving the trajectory optimization problem for the whole maneuver, and not just for the current MPC horizon. In this setting, the final state constraints in \cref{eq:boundary_constraints_SOCP_umin} should be anyway softened to address this tracking objective.
Before the softening is formally introduced, other constraints are identified to be potential sources of infeasibility. Namely, the collision avoidance constraints. These constraints could induce infeasibility in a variety of ways, one of which is when a satellite is initially situated inside the KOZ of another satellite, which may happen during operation of the closed control loop due to a variety of reasons including unmodeled nonlinearities, navigation inaccuracies, or actuation errors. Indeed, the initial state is not controlled, and if the initial positions of the satellites are not collision-free, the whole guidance problem will be doomed infeasible. It is for this reason that the guaranteed feasible trajectory optimization problem will not constrain the first step to be collision-free. Another way the collision avoidance constraints could induce infeasibility is, again, when the initial states are not collision-free, and the maximum control acceleration is not enough to get the satellite that violates the collision avoidance criteria out of the KOZ within the given time period before the following one or more time steps. To address these concerns, the collision avoidance constraint is also softened for all the time steps to ensure feasibility.\\

The SOCP formulation, including the minimum acceleration constraints, is softened and is rewritten as,
\begin{Problem}[Softened SOCP formulation]
\label{prob:SOCP_formulation_soft}
\begin{align}
& \begin{multlined}[t][\multlinedwidth]
    \min_{\mat{Y}, \overline{\mat{U}}, \mat{\Gamma}, w, \dist{B}, \Upsilon} \quad \frac{1}{a_{c}}\sum_{i\in \dist{I}}\sum_{k \in \dist{K}_{f} \setminus \dist{K}_{i}}{\parenth{\Delta t_{k} \Gamma_{i, k}}} +\\
    w + q_\text{umin} \sum_{i\in \dist{I}}\sum_{k \in \dist{K}_{f} \setminus \dist{K}_{i}}{\upsilon_{i, k}} + \\
    q_\text{ca} \sum_{i \in \dist{I}} \sum_{\substack{j \in \dist{I}\setminus \curlyb{i}\\ \cup \curlyb{0}}} \sum_{k \in \dist{K} \setminus \curlyb{0}}{\beta_{ij, k}} 
    \end{multlined} \nonumber\\
& \text{subject to,} \nonumber\\
& \vec{y}_{i, 0} = \vec{y}_{i, 0} \quad \forall i \in \dist{I}, \label{eq:boundary_constraints_SOCP_soft}\\
& \norm{\brack{\brack{\sqrt{\mat{Q}} \parenth{\vec{y}_{i, k} - \overline{\vec{y}}_{i, k}}}^{h}_{i\in \dist{I}}}^{v}_{k \in \overline{\dist{K}}}}[F] \leq w, \label{eq:soft_Yf_constraint_SOCP_soft}\\
& \vec{y}_{i, k+1} =  \mat{\Phi}_{k} \vec{y}_{i, k} + \mat{\Psi}_{k} \overline{\vec{u}}_{i, k} \quad \forall i \in \dist{I} ,\; \forall k \in \dist{K},\label{eq:dynamics_constraint_SOCP_soft}\\
& \overline{\vec{u}}_{i, k} = \vec{0} \quad \forall i \in \dist{I},\; \forall \parenth{i, k} \in \parenth{i, \dist{K}_{n} \cup \dist{K}_{i}},  \label{eq:u0_constraint_SOCP_soft}\\
& \begin{multlined}[t][\multlinedwidth]
    \frac{\breve{\overline{\vec{u}}}_{i, k}^{\intercal}}{\norm{\breve{\overline{\vec{u}}}_{i, k}}} \overline{\vec{u}}_{i, k} \geq a_{c} u_{i, \text{min}} - \upsilon_{i,k} \\
    \forall i \in \dist{I} ,\; \forall \parenth{i, k} \in \parenth{i, \dist{K}_{f} \setminus \dist{K}_{i}},
    \end{multlined}\label{eq:umin_constraint_SOCP_soft}\\
& 0 \leq \upsilon_{i,k} \leq \upsilon_\text{max} \quad
\forall i \in \dist{I} ,\; \forall \parenth{i, k} \in \parenth{i, \dist{K}_{f} \setminus \dist{K}_{i}},\\
& \norm{\overline{\vec{u}}_{i, k}}[\mat{R}] \leq \Gamma_{i,k} \quad \forall i \in \dist{I} ,\; \forall \parenth{i, k} \in \parenth{i, \dist{K}_{f} \setminus \dist{K}_{i}}, \label{eq:umax_constraint_SOCP_soft}\\
& \Gamma_{i,k} \leq a_{c} u_{i, \text{max}} \quad \forall i \in \dist{I},\; \forall \parenth{i, k} \in \parenth{i, \dist{K}_{f} \setminus \dist{K}_{i}},\\
& 
\begin{multlined}[t][\multlinedwidth]
    \dfrac{\parenth{\breve{\vec{y}}_{i, k} - \breve{\vec{y}}_{j, k}}^{\intercal}}{\norm{\mat{T}_{k}\parenth{\breve{\vec{y}}_{i, k} - \breve{\vec{y}}_{j, k}}}} \mat{T}_{k}^{\intercal} \mat{T}_{k} \parenth{\vec{y}_{i, k} - \vec{y}_{j, k}} \geq R_\text{CA} - \beta_{ij, k} \\
    \quad \forall i, j \in \dist{I} ,\; i \neq j ,\; \forall k \in \dist{K} \setminus \curlyb{0},
\end{multlined}
\label{eq:CA_deputy_deputy_SOCP_soft}\\
& \begin{multlined}[t][\multlinedwidth]
    \dfrac{\breve{\vec{y}}_{i, k}^{\intercal}} {\norm{\mat{T}_{k} \breve{\vec{y}}_{i, k}}} \mat{T}_{k}^{\intercal} \mat{T}_{k} \vec{y}_{i, k} \geq R_\text{CA} - \beta_{i0,k} \\
    \forall i \in \dist{I} ,\; \forall k \in \dist{K} \setminus \curlyb{0},
\end{multlined}\label{eq:CA_deputy_chief_SOCP_soft}\\
& \begin{multlined}[t][\multlinedwidth]
    0 \leq \beta_{ij, k} \leq \beta_\text{max} \\
    \forall i \in \dist{I}, \forall j \in \dist{I} \cup \curlyb{0}, i\neq j, \forall k \in \dist{K} \setminus \curlyb{0},
\end{multlined}\label{eq:beta_max_SOCP_soft}
\end{align}
\end{Problem}
where $w$ is a slack variable, $q_\text{umin} \in \set{R}^{+}$ and $q_\text{ca} \in \set{R}^{+}$ are positive weighting factors, $\upsilon_{i,k}$ are slack variables that represent violations to the relaxed minimum acceleration constraints, and $\beta_{ij,k}$ and $\beta_{i0,k}$ are slack variables that represent violations to the affine collision avoidance constraints. These slack variables are collated in the sets $\Upsilon$ and  $\dist{B}$, respectively, which are defined as,
\begin{equation}
    \begin{split}
    \Upsilon & = \curlyb{\upsilon_{i,k}: i \in \dist{I}, \parenth{i, k} \in \parenth{i, \dist{K}_{f} \setminus \dist{K}_{i}}},\\
    \dist{B} & = \curlyb{\beta_{ij,k}: i \in \dist{I}, j \in \dist{I} \cup \curlyb{0}, i\neq j, k \in \dist{K} \setminus \curlyb{0}}.
    \end{split}
\end{equation}
Moreover, $\overline{\dist{K}}$ is the set of indices for the time instances on which the user requires the state of the deputies to track certain reference states,
$\overline{\vec{y}}_{i, k} \in \set{R}^{6}$ is the user-defined reference state required to be tracked by the $i^\text{th}$ deputy at the time step $t_{k}$, $\norm{\cdot}[F]$ is the Frobenius norm  operator,
$\mat{Q} \in \set{R}^{6\times 6}$ and $\mat{R} \in \set{R}^{3\times 3}$ are diagonal positive semi-definite weighting matrices relating respectively to the minimization of the state deviation from the reference one and to the minimization of the total control effort (Delta-V). Note that $\norm{\overline{\vec{u}}_{i, k}}[\mat{R}] = \norm{\sqrt{\mat{R}}\overline{\vec{u}}_{i, k}}$, where the diagonal entries of $\mat{R}$ have to be strictly greater than or equal to unity, so that the acceleration could never exceed the maximum allowable one. Furthermore, since both $\mathbf{Q}$ and $\mathbf{R}$ are diagonal matrices, their matrix square roots are equal to their element-wise square roots.\\

\begin{figure}[ht]
    \centering
    \includegraphics[width=\columnwidth]{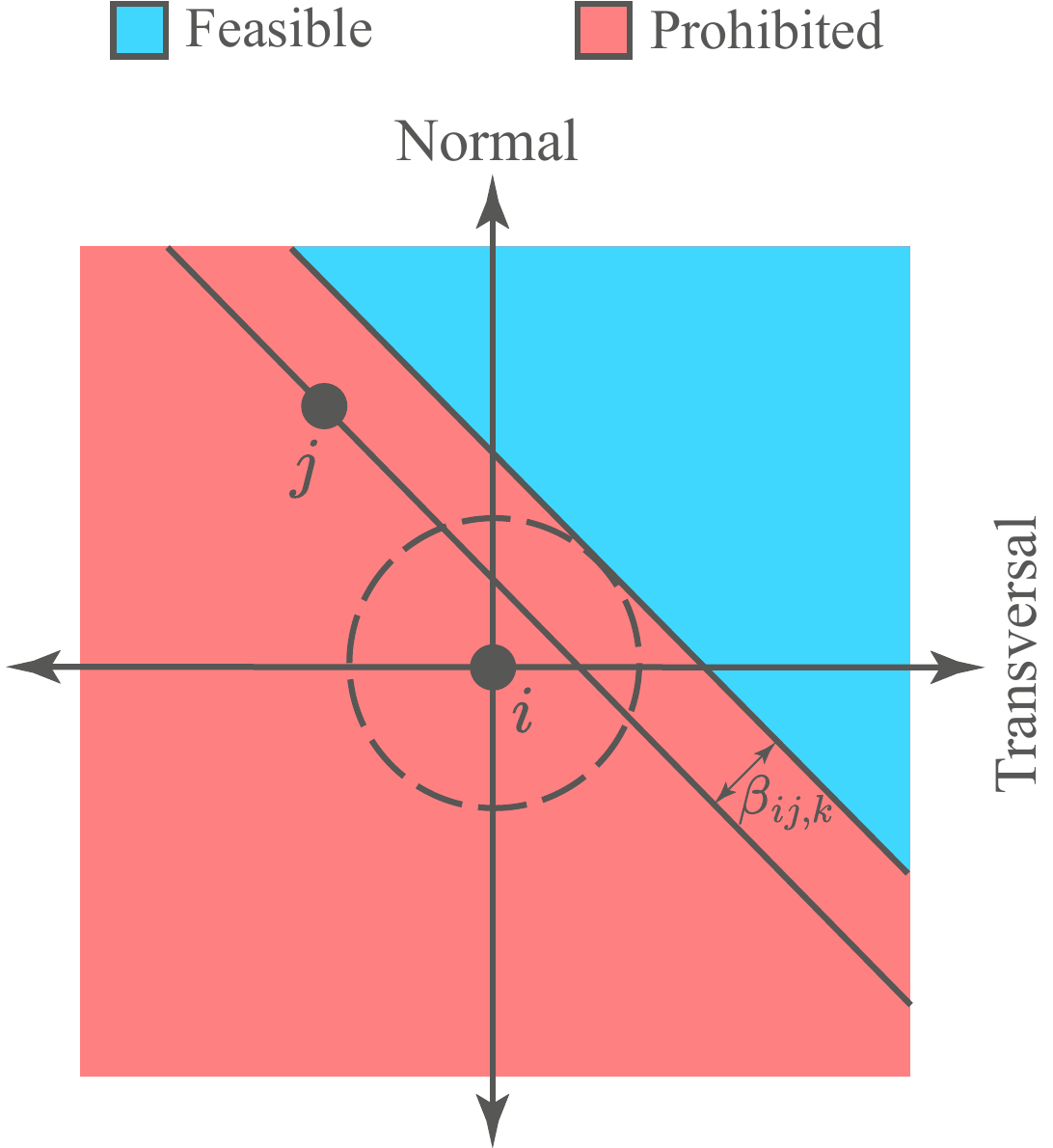}
    \caption{Violation of the affine relaxation of the collision avoidance constraint}
    \label{fig:violation_of_CA_constraints}
\end{figure}

Although the set $\dist{B}$ collates the values corresponding to violations in the affine relaxation of the collision avoidance constraints, non-zero values of the elements of $\dist{B}$ do not necessarily indicate violations to the original collision avoidance constraints, i.e., keeping one satellite outside the KOZ of another. This concept is illustrated visually in \cref{fig:violation_of_CA_constraints} (assuming the motion is taking place in 2D plane, without loss of generality) where it can be seen that while the affine relaxation of the collision avoidance constraint is allowed to be violated, the original constraint is still respected. The same can be said about violations to the relaxed minimum acceleration constraints. A non-zero $\upsilon_{i, k}$ does not necessarily imply that the control acceleration required by the $i^\text{th}$ deputy at time $t_{k}$ is below the minimum acceleration threshold. It means, however, that one of the relaxed minimum acceleration constraint is violated, which might or might not lead to a violation in the original constraint, since the relaxation is overly restrictive.

Problem \ref{prob:SOCP_formulation_soft} is written in the epigraph form, which behaves such that,
\begin{equation}
\begin{split}
    w & = \norm{\brack{\brack{\sqrt{\mat{Q}} \parenth{\vec{y}_{i, k} - \overline{\vec{y}}_{i, k}}}^{h}_{i\in \dist{I}}}^{v}_{k \in \overline{\dist{K}}}}[F],\\
    & = \sqrt{\sum_{i\in \dist{I}} \sum_{k \in \overline{\dist{K}}}{\parenth{\parenth{\vec{y}_{i, k} - \overline{\vec{y}}_{i, k}}^{\intercal} \mat{Q} \parenth{\vec{y}_{i, k} - \overline{\vec{y}}_{i, k}}}}}.
\end{split}
\end{equation}

Note that Problem \ref{prob:SOCP_formulation_soft} reduces to a softened version of Problem \ref{prob:SOCP_formulation_umin} if and only if $\overline{\dist{K}} = \{m+1\}$ and $\overline{\vec{y}}_{i, m+1} = \overline{\vec{y}}_{i, f}$. Otherwise, the two problems become unrelated. It is worth noting that the matrices $\mat{R}$ and $\mat{Q}$ can be adjusted to obtain specific desired solutions without changing the structure of the problem. These desired solutions may include ones that avoid using the radial acceleration component or those that track a given ROE profile, excluding, for example, the relative mean argument of latitude. A third example is when the operator requires more emphasis on minimizing the errors of specific ROE elements, such as the relative semi-major axis, for being more critical than others. \\

In order to showcase the importance of softening the hard constraints, the SOCP guidance problem in their two settings, i.e., the hard- and the soft-constrained ones, in Problems \ref{prob:SOCP_formulation_umin} and \ref{prob:SOCP_formulation_soft} respectively, are run over two of the reconfiguration scenarios for comparison purposes. The initial and final states for the two reconfiguration scenarios are provided in \ref{app:Reconfiguration_scenarios}, with only the first and second reconfigurations being used in this comparison. The selected configurations for these scenarios are chosen for their relevance to real-world applications. The characteristics and significance of these configurations are thoroughly discussed in \cite{Fasano2014Formation_Geometry, Lopez-Dekker2013Formation_geometry, Wang2022Optimal_PCO}, where several of them are identified as commonly used in multi-static SAR missions.
The deputy spacecraft are assumed identical in both scenarios, and hence $u_{i, \text{max}} = u_{\text{max}},\; u_{i, \text{min}} = u_{\text{min}},\; p_{i} = p \; \forall i \in \dist{I}$. Furthermore, since the maneuver duration, $t_{f} - t_{0}$, is defined for the two scenarios, the user-defined time vector, $\vec{t}$, is provided for both maneuvers such that the duration of the thrust arcs are identical throughout the maneuver. This also applies to the durations of the coast arcs. Formally, $T_{f,l} = T_{f},\; T_{n,l} = T_{n}, \; \forall l \in \dist{L}$, where $\dist{L} = \curlyb{1, 2, \hdots, \parenth{m+1}/2}$ is the list of indices of the different control cycles (refer to \cref{fig:Low-thrust-guidance-scheme}). It is important to emphasize that 
the soft-constrained problem is solved with $\overline{\dist{K}} = \{m+1\}$ and with $\overline{\vec{y}}_{i, m+1} = \overline{\vec{y}}_{i, f} \; \forall i \in \dist{I}$. A full list of the parameters used in solving the guidance problems for both scenarios are reported in \cref{tab:guidance_soft_hard_comparison_parameters}, in which $\mat{I}_{n}$ is the identity matrix with dimensions $n\times n$. The choice of $T_{f}$ and $T_{n}$ is based on the results of the sensitivity analysis performed in \cite{mahfouz2024Delta_V_Optimal}.

{\renewcommand{\arraystretch}{\arraystretchfortable}
\begin{table}[ht]
    \centering
    \caption{Parameters used in the comparison experiment between the soft- and hard-constrained guidance problems}
    \begin{tabular}{l c c l c}
        \hline
        \hline
        Parameter & Value & ~ & Parameter & Value \\
        \hline
        $T_{f}$ [orbits] & $0.2$ & ~ & $T_{n}$ [s] & $100$ \\
        $u_\text{max}$ [\textmu m/s$^{2}$] & $35$ & ~ & $u_\text{min}$ [\textmu m/s$^{2}$] & $20$ \\
        $R_\text{CA}$ [m] & $100$ & ~ & $p$ [\%] & $100$ \\
        $\mat{Q}$ [-] & $\mat{I}_{3}$ & ~ & $\mat{R}$ [-] & $\mat{I}_{6}$ \\
        $q_\text{umin}$ [-] & $10^{-2}$ & ~ & $\upsilon_\text{max}$ [m$^{2}$/s$^{2}$] & $\infty$ \\
        $q_\text{ca}$ [-] & $1$ & ~ & $\beta_\text{max}$ [m] & $10$ \\
        \hline
        \hline
    \end{tabular}
    \label{tab:guidance_soft_hard_comparison_parameters}
\end{table}
}

The first reconfiguration scenario was chosen to showcase the significance of transforming Problem \ref{prob:SOCP_formulation_umin} into Problem \ref{prob:SOCP_formulation_soft}. Specifically, when Problem \ref{prob:SOCP_formulation_umin} is solved according to the logic of \cref{fig:SCP_scheme_umin} over Reconfiguration 1, it fails to produce a feasible solution, while it can be solved with a tolerable level of constraint violations when the formulation of Problem \ref{prob:SOCP_formulation_soft} is used. This makes the first reconfiguration scenario an excellent example of how the softened formulation is beneficial. The second reconfiguration scenario, on the other hand, can be solved by both, the hard-constrained formulation as well as the soft-constrained one, and hence forms a basis for comparing the two methodologies. The results of the comparison experiment are summarized in \cref{tab:centralized_hard_soft_comparison_results_Tc_0.2} where the number of variables, the number of constraints, the total required Delta-V for the maneuver, the maximum final state error, and the maximum violation of the relaxed minimum acceleration constraints as well as that of the relaxed collision avoidance constraints across all time steps and for all satellites are reported. In the table, a column of only dashes (i.e., -) indicates that no solution could be found for the adopted formulation over a certain reconfiguration scenario.
{\renewcommand{\arraystretch}{\arraystretchfortable}
\begin{table}[ht]
    \centering
    \caption{Comparison of the soft- and the hard-constrained problems}
    \label{tab:centralized_hard_soft_comparison_results_Tc_0.2}
    \scriptsize
    \begin{tabular}{lccccc}
    \hline
    \hline 
    ~ & \multicolumn{2}{c}{Reconfiguration 1} & ~ & \multicolumn{2}{c}{Reconfiguration 2}\\
    \cline{2-3} \cline{5-6}
    ~ & Hard & Soft & ~ & Hard & Soft\\ 
    \hline 
    Variables & - & 3547 & ~ & 1696 & 2185 \\ 
    Constraints & - & 2719 & ~ & 1632 & 1633 \\
    $\Delta V$ [m/s] & - & 2.77 & ~ & 1.69 & 1.69 \\ 
    $\text{max}\parenth{\norm{\vec{y}_{i,f} - \overline{\vec{y}}_{i,f}}}$ [m] & - & 0.00 & ~ & 0.00 & 0.00 \\ 
    $\text{max}\parenth{\Upsilon}$ [$\text{m}^{2}/\text{s}^{2}$] & - & 0.0 & ~ & - & 0.0 \\ 
    $\text{max}\parenth{\dist{B}}$ [m] & - & 0.0 & ~ & - & 0.0 \\  
    \hline 
    \hline 
    \end{tabular}
\end{table}
}

The results in \cref{tab:centralized_hard_soft_comparison_results_Tc_0.2} indicate that the hard-constrained problem formulation did indeed fail to produce a feasible solution for the first reconfiguration scenario. The effectiveness of introducing slack variables becomes particularly evident in Reconfiguration $1$, as the softened problem successfully identified a solution that satisfied all constraints, including the final state, collision avoidance, and minimum acceleration requirements. This suggests that the hard-constrained problem for Reconfiguration $1$ was "almost" feasible, with only a minimal constraint violation needed to yield a valid solution.
While it might be possible for the hard-constrained formulation to find a feasible solution for Reconfiguration $1$ by adjusting parameters such as the pruning factor, such an approach is unsuitable for autonomous implementation due to its reliance on manual tuning.\\

The results of Reconfiguration $2$ are interesting since the two variants of the problems, i.e., soft- and hard-constrained, were able to provide feasible solutions for it.
Generally, the results from a soft-constrained problem are close to those of the hard-constrained one, but they are not necessarily identical. This is because, as shown in \cref{fig:SCP_scheme_umin}, a problem incorporating both collision avoidance and minimum acceleration constraints is typically solved in four iterations. The solution of each iteration, except the zeroth, depends on the solution from the previous iteration. Consequently, the "final result" of interest is the outcome of the last iteration, which relies on the preceding solutions. Given the complexity of the guidance problem, it is not guaranteed that the solutions from each iteration in the soft- and hard-constrained settings will match. Nevertheless, it can be observed that the two variants produce identical solutions for Reconfiguration $2$, although this may not be the case for a general reconfiguration scenario. \\

The comparison between the soft- and hard-constrained problems, as presented in \cref{tab:centralized_hard_soft_comparison_results_Tc_0.2}, was designed to showcase the softened guidance scheme's capability to handle infeasibility. This issue is particularly relevant in closed-loop operations where guidance strategies are repeatedly employed, as discussed in \cref{sec:Control}. One of the key challenges in this context is ensuring collision-free trajectories throughout the maneuver. The guidance strategies developed so far guarantee collision avoidance only at discrete sampling times, $t_{k} \; \forall k \in \dist{K}$, leaving the safety of the relative trajectories between these sampling times uncertain.
Using a control cycle duration of $T_{f} = 0.2$ orbits poses a potential risk for closed-loop implementation. As the interval between successive sampling times increases, the probability of a satellite entering the Keep-Out Zone (KOZ) of another satellite becomes significantly higher. To address this issue, the thrust arc length within each control cycle is reduced here, and in the discussions to follow, to $T_{f} = 0.05$ orbits, roughly equivalent to 300 seconds. This adjustment ensures finer control authority and minimizes the risk of collisions during the unmonitored intervals between sampling times.
To evaluate the impact of this refined setting on the guidance layer, the comparison experiment between the hard- and soft-constrained guidance problems was repeated. The results, summarized in \cref{tab:centralized_hard_soft_comparison_results_Tc_0.05}, reveal a notable increase in the number of variables and constraints, thereby significantly raising the computational complexity. Despite this added complexity, the shorter thrust arcs allowed the hard-constrained problem to achieve a feasible solution for Reconfiguration $1$, demonstrating the benefits of enhanced control precision in this configuration.
{\renewcommand{\arraystretch}{\arraystretchfortable}
\begin{table}[ht]
    \centering
    \caption{Comparison of the soft- and the hard-constrained problems}
    \label{tab:centralized_hard_soft_comparison_results_Tc_0.05}
    \scriptsize
    \begin{tabular}{lccccc}
    \hline
    \hline 
    ~ & \multicolumn{2}{c}{Reconfiguration 1} & ~ & \multicolumn{2}{c}{Reconfiguration 2}\\
    \cline{2-3} \cline{5-6}
    ~ & Hard & Soft & ~ & Hard & Soft\\ 
    \hline 
    Variables & 8244 & 11554 & ~ & 5496 & 7115 \\ 
    Constraints & 8925 & 8926 & ~ & 5362 & 5363 \\ 
    $\Delta V$ [m/s] & 2.58 & 2.68 & ~ & 1.58 & 1.58 \\ 
    $\text{max}\parenth{\norm{\vec{y}_{i,f} - \overline{\vec{y}}_{i,f}}}$ [m] & 0.00 & 0.00 & ~ & 0.00 & 0.00 \\ 
    $\text{max}\parenth{\Upsilon}$ [$\text{m}^{2}/\text{s}^{2}$] & - & 0.0 & ~ & - & 0.0 \\ 
    $\text{max}\parenth{\dist{B}}$ [m] & - & 0.0 & ~ & - & 0.0 \\ 
    \hline 
    \hline 
    \end{tabular}
\end{table}
}


\subsection{Distributed guidance}
Up to this point, all the presented guidance schemes have relied on a centralized processing unit, namely the chief spacecraft, to compute the guidance profiles. However, this centralized approach is not scalable, as the computational time required to solve the problem increases exponentially with the number of deputies, as demonstrated in \cite{mahfouz2024Delta_V_Optimal}. To address this limitation, the guidance problem must be reformulated to allow parallel computation by each deputy spacecraft.\\

A closer inspection of Problem \ref{prob:SOCP_formulation_soft} reveals that the primary obstacle preventing a distributed solution lies in the constraints in \cref{eq:CA_deputy_deputy_SOCP_soft}. Specifically, this issue arises because the state profile of the $j^\text{th}$ deputy at the current SCP iteration, ${\vec{y}}_{j, k}$, is not directly accessible onboard the $i^\text{th}$ satellite. To resolve this, the state profile of the $j^\text{th}$ deputy from the previous SCP iteration, $\breve{\vec{y}}_{j, k}$, can be used instead. With this modification, Problem \ref{prob:SOCP_formulation_soft} can be reformulated into a distributed framework and can be rewritten for the $i^\text{th}$ deputy as follows:
\begin{Problem}[Softened distributed SOCP formulation]
\label{prob:SOCP_formulation_soft_dist}
\begin{align}
& \begin{multlined}[t][\multlinedwidth]
    \min_{\mat{Y}_{i}, \overline{\mat{U}}_{i}, \mat{\Gamma}_{i}, w, \dist{B}_{i}, \Upsilon_{i}} \quad \frac{1}{a_{c}}\sum_{k \in \dist{K}_{f} \setminus \dist{K}_{i}}{\parenth{\Delta t_{k} \Gamma_{i, k}}} + w +\\
    q_\text{umin} \sum_{k \in \dist{K}_{f} \setminus \dist{K}_{i}}{\upsilon_{i, k}} + 
    q_\text{ca} \sum_{\substack{j \in \dist{I}\setminus \curlyb{i}}} \sum_{k \in \dist{K} \setminus \curlyb{0}}{\beta_{ij, k}} 
    \end{multlined} \nonumber\\
& \text{subject to,} \nonumber\\
& \vec{y}_{i, 0} = \vec{y}_{i, 0}, \label{eq:boundary_constraints_SOCP_soft_dist}\\
& \norm{\brack{\sqrt{\mat{Q}} \parenth{\vec{y}_{i, k} - \overline{\vec{y}}_{i, k}}}^{v}_{k \in \overline{\dist{K}}}} \leq w, \label{eq:soft_Yf_constraint_SOCP_soft_dist}\\
& \vec{y}_{i, k+1} =  \mat{\Phi}_{k} \vec{y}_{i, k} + \mat{\Psi}_{k} \overline{\vec{u}}_{i, k} \quad \forall k \in \dist{K},\label{eq:dynamics_constraint_SOCP_soft_dist}\\
& \overline{\vec{u}}_{i, k} = \vec{0} \quad \forall \parenth{i, k} \in \parenth{i, \dist{K}_{n} \cup \dist{K}_{i}},  \label{eq:u0_constraint_SOCP_soft_dist}\\
& \begin{multlined}[t][\multlinedwidth]
    \frac{\breve{\overline{\vec{u}}}_{i, k}^{\intercal}}{\norm{\breve{\overline{\vec{u}}}_{i, k}}} \overline{\vec{u}}_{i, k} \geq a_{c} u_{i, \text{min}} - \upsilon_{i,k} \\
    \forall i \in \dist{I} ,\; \forall \parenth{i, k} \in \parenth{i, \dist{K}_{f} \setminus \dist{K}_{i}},
    \end{multlined}\label{eq:umin_constraint_SOCP_soft_dist}\\
& 0 \leq \upsilon_{i,k} \leq \upsilon_\text{max} \quad
\forall \parenth{i, k} \in \parenth{i, \dist{K}_{f} \setminus \dist{K}_{i}},\\
& \norm{\overline{\vec{u}}_{i, k}}[\mat{R}] \leq \Gamma_{i,k} \quad \forall i \in \dist{I} ,\; \forall \parenth{i, k} \in \parenth{i, \dist{K}_{f} \setminus \dist{K}_{i}}, \label{eq:umax_constraint_SOCP_soft_dist}\\
& \Gamma_{i,k} \leq a_{c} u_{i, \text{max}} \quad \forall \parenth{i, k} \in \parenth{i, \dist{K}_{f} \setminus \dist{K}_{i}},\\
& 
\begin{multlined}[t][\multlinedwidth]
    \dfrac{\parenth{\breve{\vec{y}}_{i, k} - \breve{\vec{y}}_{j, k}}^{\intercal}}{\norm{\mat{T}_{k}\parenth{\breve{\vec{y}}_{i, k} - \breve{\vec{y}}_{j, k}}}} \mat{T}_{k}^{\intercal} \mat{T}_{k} \parenth{\vec{y}_{i, k} - \breve{\vec{y}}_{j, k}} \geq R_\text{CA} - \beta_{ij, k} \\
    \quad \forall j \in \dist{I}\setminus{\curlyb{i}},\; \forall k \in \dist{K} \setminus \curlyb{0},
\end{multlined}
\label{eq:CA_deputy_deputy_SOCP_soft_dist}\\
& \begin{multlined}[t][\multlinedwidth]
    0 \leq \beta_{ij, k} \leq \beta_\text{max} \quad 
    \forall j \in \dist{I} \setminus \curlyb{i}, \forall k \in \dist{K} \setminus \curlyb{0},
\end{multlined}\label{eq:beta_max_SOCP_soft_dist}
\end{align}
\end{Problem}
where,
\begin{equation}
\begin{split}
    \mat{Y}_{i} & = \brack{\vec{y}_{i, k}}^{h}_{k \in \dist{K} \cup \curlyb{m+1}},\\  &\\ 
    \overline{\mat{U}} & = \brack{\overline{\vec{u}}_{i, k}}^{h}_{k \in \dist{K}},\\
    &\\
    \mat{\Gamma}_{i} & = \brack{\Gamma_{i, k}}^{h}_{k \in \dist{K}_{f}},\\
    \Upsilon_{i} & = \curlyb{\upsilon_{i,k}: \parenth{i, k} \in \parenth{i, \dist{K}_{f} \setminus \dist{K}_{i}}},\\
    \dist{B}_{i} & = \curlyb{\beta_{ij,k}: j \in \dist{I} \setminus \curlyb{i}, k \in \dist{K} \setminus \curlyb{0}}.
\end{split}
\end{equation}

It is important to highlight that the chief-deputy collision avoidance constraints were excluded from the distributed guidance strategy. This is because the distributed approach treats the chief spacecraft as a virtual point, eliminating the need for a central processing unit since guidance calculations are performed onboard each deputy. Despite being considered a virtual point, the chief is assumed to orbit the Earth considering the effect of the second zonal harmonic ($J_2$).\\

\begin{figure*}[ht]
    \centering
    \includegraphics[width=\linewidth]{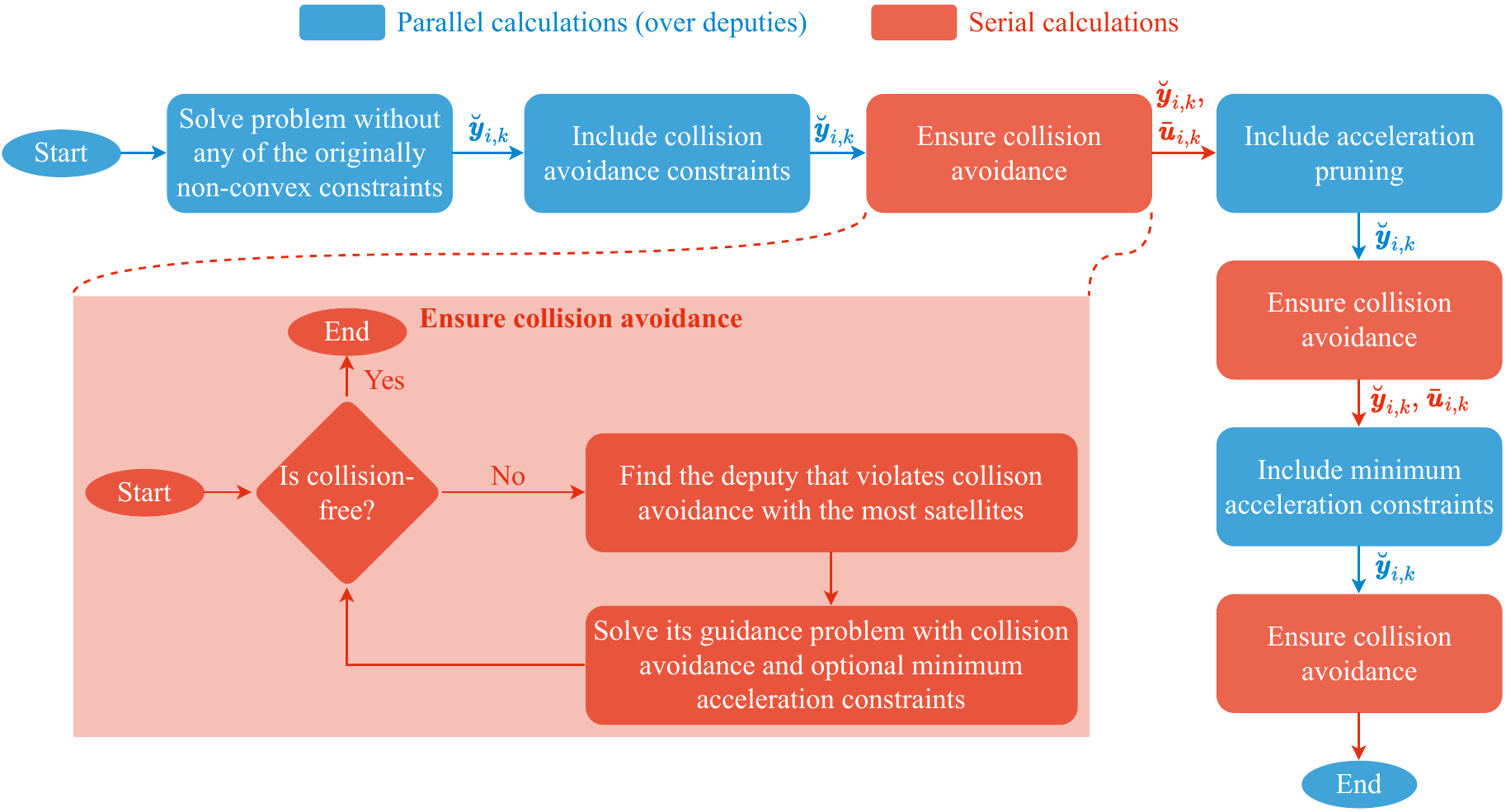}
    \caption{Graphical representation of the SCP scheme in the distributed setting}
    \label{fig:SCP_scheme_umin_dist}
\end{figure*}

Similar to Problem \ref{prob:SOCP_formulation_soft}, Problem \ref{prob:SOCP_formulation_soft_dist} is solved using a Sequential Convex Programming (SCP) scheme, where most of the convex sub-problems are solved simultaneously for all deputies. However, not all sub-problems can be solved in parallel due to the collision avoidance constraints. Each deputy is only aware of the optimized trajectories of the other deputies from the previous SCP step and not the current one. As a result, the computed trajectories are not guaranteed to be collision-free, even after multiple iterations.
If all calculations are performed in parallel, a "ping-pong" situation may arise. To explain, consider Satellites A and B solving their trajectory optimization problems individually and in parallel. The resulting trajectories may be unsafe. If they re-solve their problems in parallel, Satellite A computes a new trajectory that is safe relative to the old trajectory of Satellite B, but not the new trajectory. Similarly, Satellite B computes a trajectory that is safe relative to the old trajectory of Satellite A. This process can repeat indefinitely, resulting in oscillations between unsafe trajectories.
To mitigate this issue, some serial computations are introduced. In this approach, all satellites temporarily fix their trajectories while one satellite optimizes its trajectory, considering the fixed trajectories of all others. This sequential process ensures that the updated trajectory is safe relative to the current trajectories of all other satellites. If the overall configuration remains unsafe, the process is repeated iteratively. The execution logic for this distributed SCP scheme is illustrated in \cref{fig:SCP_scheme_umin_dist}, where the mechanism ensuring collision-free final guidance profiles is also detailed.

Clearly, the "Ensure collision avoidance" block in \cref{fig:SCP_scheme_umin_dist} addresses the challenge of avoiding the ping-pong effect by properly scheduling the calculations for each satellite to guarantee safe relative trajectories. Placing this block as the final step in the execution logic is natural, as the final trajectories must be collision-free. However, this block is positioned after the collision avoidance and pruning blocks to improve the overall problem's feasibility.\\

It is worth mentioning that while Problem \ref{prob:SOCP_formulation_soft_dist} incorporates the collision avoidance, minimum acceleration, and final state constraints in their softened forms, there is a corresponding formulation where these constraints are strictly hard. However, this hard-constrained version is not explicitly discussed here for the sake of brevity and is left as an exercise for the reader. To ensure consistency with the centralized guidance schemes, the experiment comparing the hard- and soft-constrained formulations, with results presented in \cref{tab:centralized_hard_soft_comparison_results_Tc_0.05}, is conducted for the distributed case as well. The outcomes of this distributed comparison are summarized in \cref{tab:distributed_hard_soft_comparison_results_Tc_0.05}.
{\renewcommand{\arraystretch}{\arraystretchfortable}
\begin{table}[ht]
    \centering
    \caption{Comparison of the soft- and the hard-constrained problems in the distributed setting}
    \label{tab:distributed_hard_soft_comparison_results_Tc_0.05}
    \scriptsize
    \begin{tabular}{lccccc}
    \hline
    \hline 
    ~ & \multicolumn{2}{c}{Reconfiguration 1} & ~ & \multicolumn{2}{c}{Reconfiguration 2}\\
    \cline{2-3} \cline{5-6}
    ~ & Hard & Soft & ~ & Hard & Soft\\ 
    \hline 
    Variables & 1374 & 2145.67 & ~ & 1374 & 1851.5 \\ 
    Constraints & 1703.5 & 1707.67 & ~ & 1412.5 & 1413.5 \\ 
    $\Delta V$ [m/s] & 2.59 & 2.76 & ~ & 1.58 & 1.58 \\ 
    $\text{max}\parenth{\norm{\vec{y}_{i,f} - \overline{\vec{y}}_{i,f}}}$ [m] & 0.00 & 0.00 & ~ & 0.00 & 0.00 \\ 
    $\text{max}\parenth{\Upsilon}$ [$\text{m}^{2}/\text{s}^{2}$] & - & 0.0 & ~ & - & 0.0 \\ 
    $\text{max}\parenth{\dist{B}}$ [m] & - & 0.0 & ~ & - & 0.0 \\
    \hline 
    \hline 
    \end{tabular}
\end{table}
}

The reported numbers of variables and constraints in \cref{tab:distributed_hard_soft_comparison_results_Tc_0.05} represent the averages across all deputies for the final SCP iteration. Notably, the average values for the two reconfigurations are generally not whole numbers, indicating that the deputies did not have identical numbers of variables and constraints at the last SCP iteration, despite solving the same guidance problem. This discrepancy arises because the cardinality of ${\dist{K}}_{i}$ varies among the deputies. In other words, differences in the number of pruned time instances for each satellite result in varying numbers of constraints. Consequently, the number of variables differs due to variations in the number of slack variables used to soften the minimum acceleration constraints, $\abs{\Upsilon_{i}}$, which might be unique to each satellite.\\

\begin{figure*}[ht]
\begin{subfigure}[c]{\columnwidth}
    \centering
    \includegraphics[width=\columnwidth]{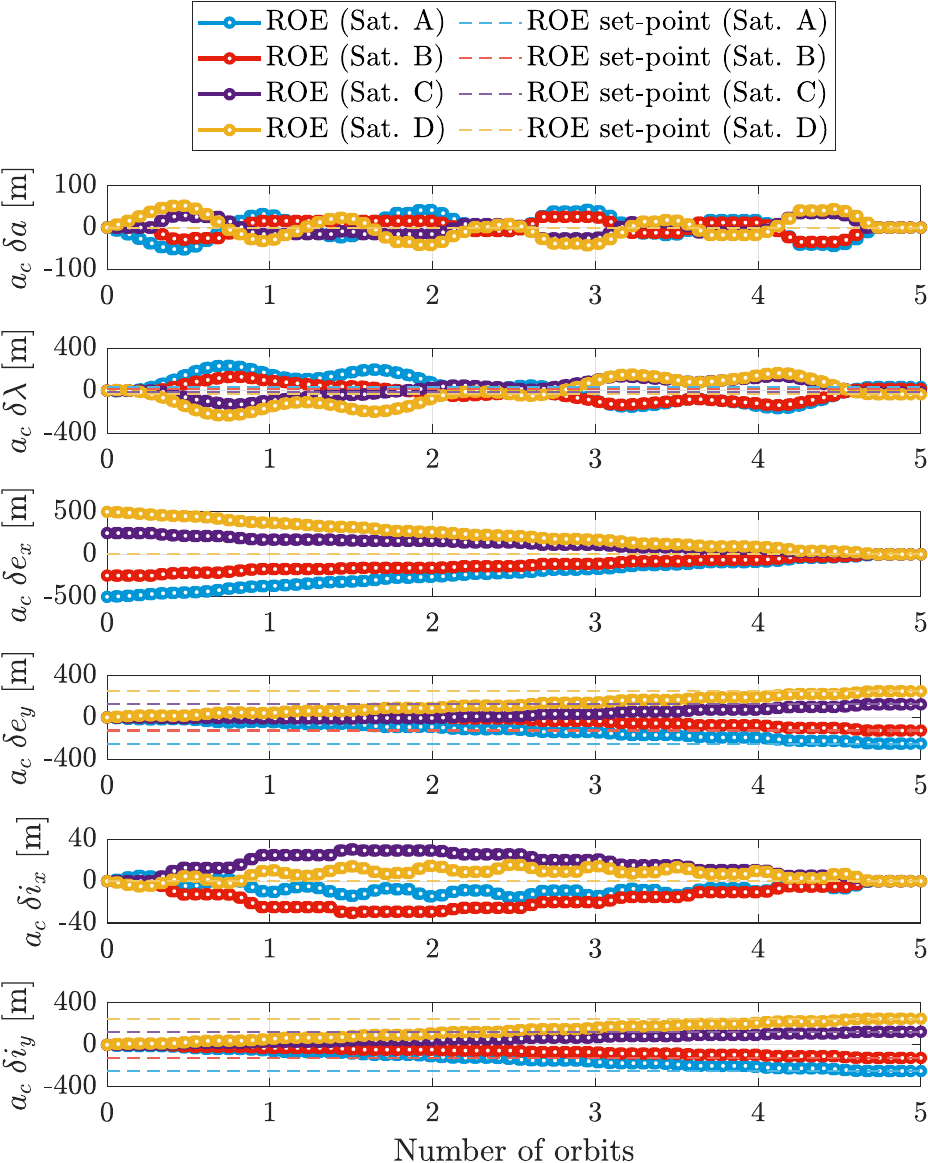}
    \caption{Solution of the hard-constrained problem} 
\end{subfigure}
\hfill
\begin{subfigure}[c]{\columnwidth}
    \centering
    \includegraphics[width=\columnwidth]{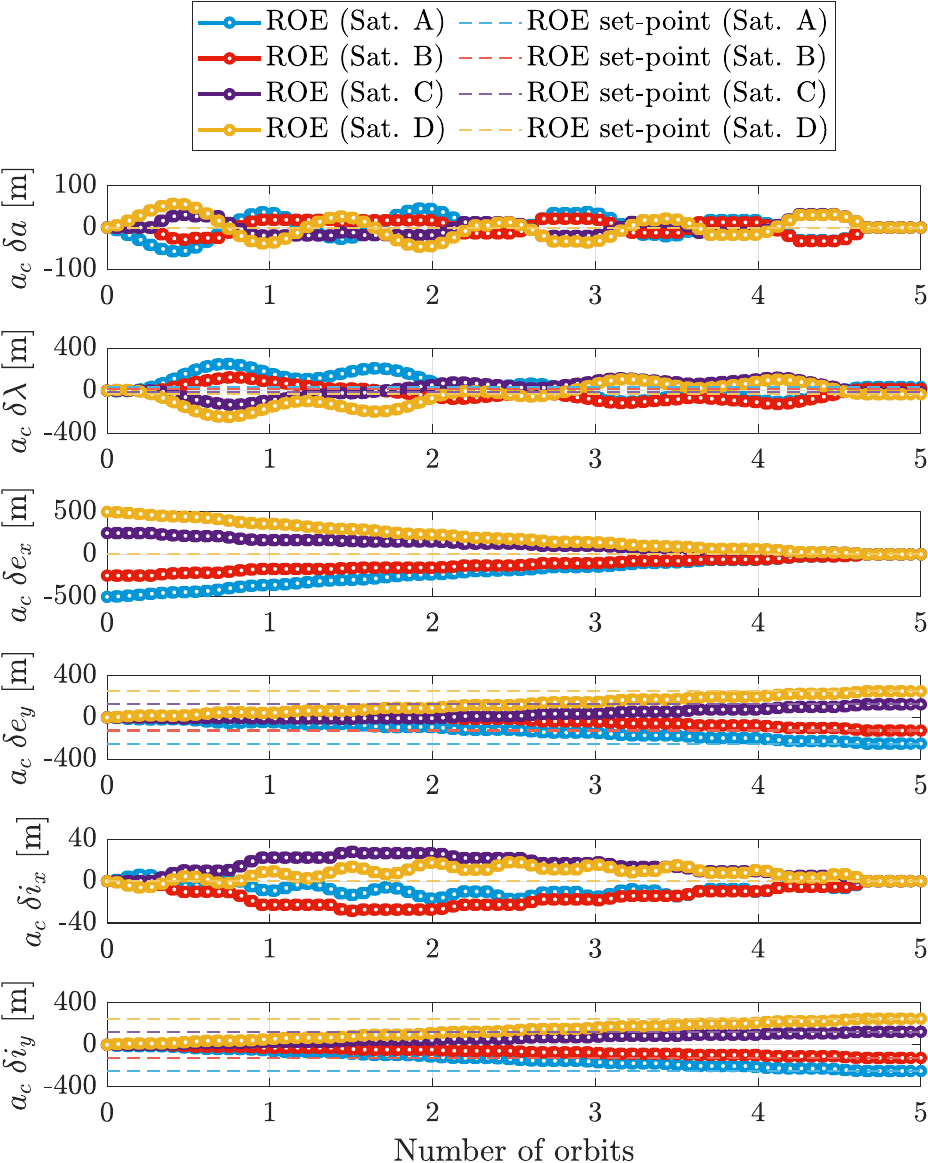}
    \caption{Solution of the soft-constrained problem}
\end{subfigure}
\caption{ROE profiles resulting from solving the distributed guidance problems over Reconfiguration $2$}
\label{fig:hard_soft_ROE_traj_dist_guidance}
\end{figure*}

The results presented in \cref{tab:distributed_hard_soft_comparison_results_Tc_0.05} highlight the significant reduction in the number of variables and constraints when compared to the centralized case (see \cref{tab:centralized_hard_soft_comparison_results_Tc_0.05}). Indeed, the number of variables and constraints increases linearly with the number of deputies, driven solely by the inclusion of collision avoidance constraints. Another notable observation, evident from the comparison of results in Tables {\ref{tab:centralized_hard_soft_comparison_results_Tc_0.05}} and {\ref{tab:distributed_hard_soft_comparison_results_Tc_0.05}}, is that the distributed guidance strategy generally requires a higher Delta-V compared to the centralized strategy. Furthermore, as claimed earlier, the softened problem generally exhibits a distinct guidance profile compared to its hard-constrained counterpart, although the Delta-V requirements for the two trajectories do not differ substantially. 
The state profiles in the ROE space, corresponding to the solutions of both the hard- and soft-constrained distributed problems for Reconfiguration $2$, are depicted in \cref{fig:hard_soft_ROE_traj_dist_guidance}, illustrating the close resemblance between the trajectories followed under the two settings.

To demonstrate the effectiveness of the distributed approach, an experiment was conducted where the number of deputy satellites varied from 1 to 20. The formation was tasked with executing a Coplanar-to-Projected Circular Orbit (PCO) maneuver. Initially, the distance between each pair of consecutive deputies was set to $200\;\text{m}$, and the radius of the final PCO was set to $500\;\text{m}$. For consistency, each of the 20 reconfigurations was allocated 10 orbits to complete.\\

\begin{figure}[ht]
    \centering
    \includegraphics[width=\linewidth]{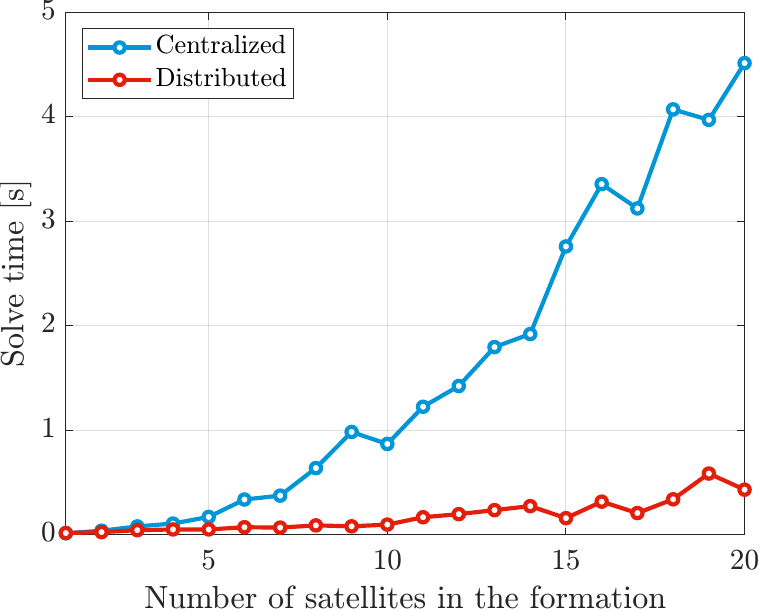}
    \caption{Solve time as the number of deputies increases}
    \label{fig:Cent_VS_Dist_solvetime}
\end{figure}

For each reconfiguration scenario, Problems {\ref{prob:SOCP_formulation_soft}} and {\ref{prob:SOCP_formulation_soft_dist}} were solved multiple times, and the average computation time was recorded. The results are presented in {\cref{fig:Cent_VS_Dist_solvetime}}, where the solve time is plotted against the number of deputies. In the distributed approach, the solve time reflects the maximum computation time across all deputies since calculations are performed in parallel.

The results show that in the centralized setting, the solve time increases exponentially with the number of deputies. In contrast, the distributed setting exhibits a much slower growth in solve time. This slower growth is primarily due to collision avoidance constraints, which require additional serial calculations in the "Ensure collision avoidance" blocks, as depicted in {\cref{fig:SCP_scheme_umin_dist}}. Without these constraints, the distributed approach would maintain nearly constant solve times regardless of the number of deputies.



\section{Control}\label{sec:Control}
The guidance schemes discussed in \cref{sec:Guidance} represent open-loop control system, which, if applied directly to the formation reconfiguration system, may not achieve the desired control objectives due to unmodeled disturbances, linearization errors, and many other factors. To address these issues, it is necessary to close the control loop using sensor measurements and navigation filters. A key component of this closed-loop system is a stepping control function, which in our case uses the guidance schemes to produce the appropriate control input based on the current state of the formation.
In this work, two distinct stepping strategies are utilized: a Shrinking-Horizon MPC (SHMPC) and a Fixed-Horizon MPC (FHMPC).\\

\begin{figure}[ht]
    \centering
    \includegraphics[width=\columnwidth]{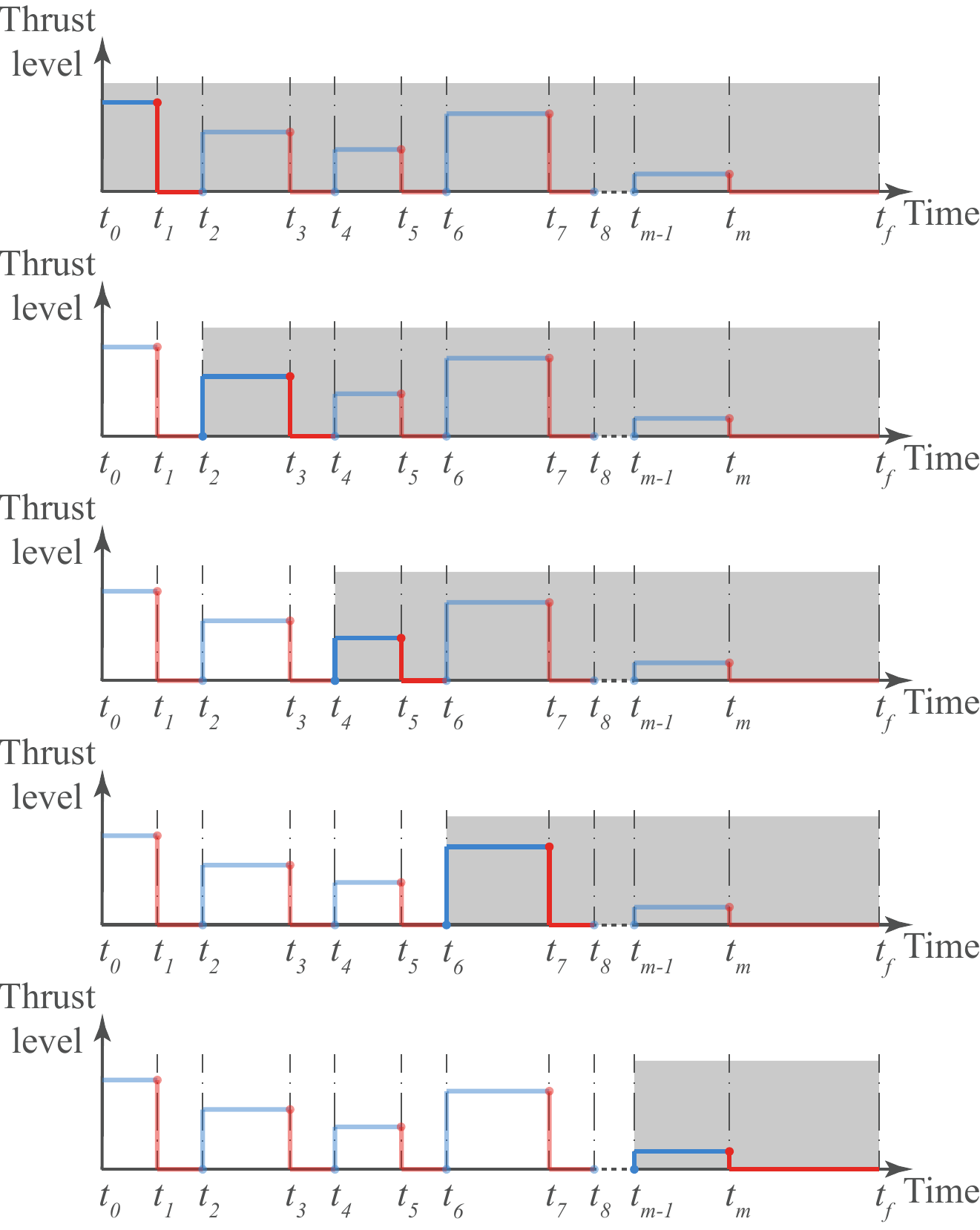}
    \caption{Evolution of the shrinking horizons}
    \label{fig:SHMPC_evolution_animation}
\end{figure}

The shrinking-horizon MPC relies on optimizing the state and the control profiles over a time span that extends from the beginning of the current control cycle to the final time of the reconfiguration. In light of \cref{fig:Low-thrust-guidance-scheme}, the SHMPC starts by solving the guidance problem over the time span from $t_{0}$ to $t_{f}$, while the second horizon extends from $t_{2}$ to $t_{f}$ and so on. The evolution of the shrinking horizons is illustrated graphically in \cref{fig:SHMPC_evolution_animation}, where the prediction horizon is shaded gray, starting at the current time $t_{k}$.

Although solving the guidance problem produces state and control profiles over the entire current horizon, only the first control cycle of each control profile is utilized and is provided as the output of the control stepping function.\\

\begin{figure}[ht]
    \centering
    \includegraphics[width=\columnwidth]{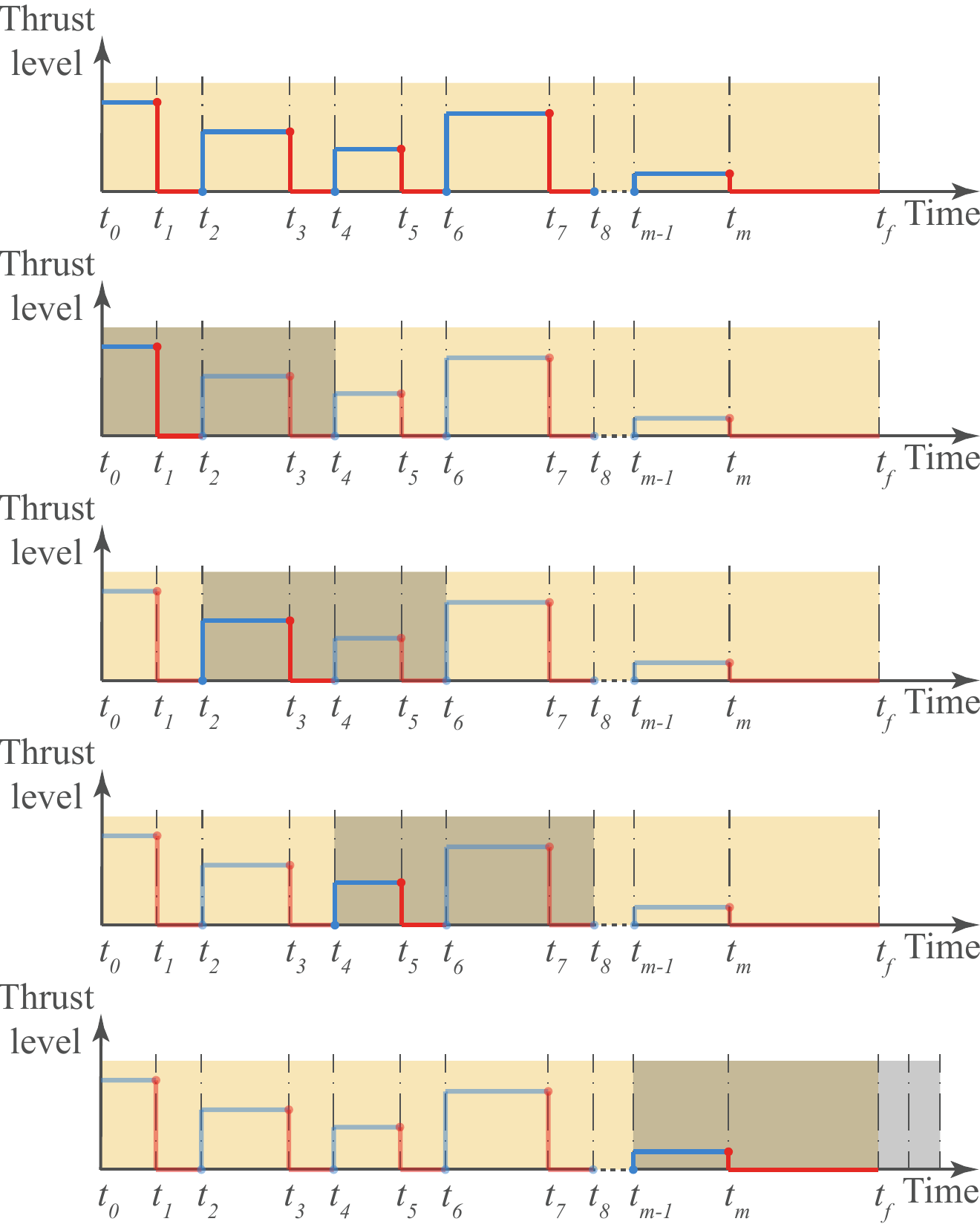}
    \caption{Evolution of the fixed horizons}
    \label{fig:FHMPC_evolution_animation}
\end{figure}

In contrast to SHMPC, FHMPC employs a fixed number of steps, $N_{h}$ with $N_{h} << m+1$, in each horizon over which Problem \ref{prob:SOCP_formulation_soft} (or Problem \ref{prob:SOCP_formulation_soft_dist}) is solved. In order for the FHMPC to operate properly, a prior optimization of the trajectory of each deputy throughout the formation reconfiguration maneuver is required, which is a one-time procedure that can be performed onboard the satellites or on the ground (see the yellow area in \cref{fig:FHMPC_evolution_animation}). The optimized trajectory is used as the reference trajectory that the guidance tries to track for each of the horizons of the FHMPC. In other words, $\overline{\dist{K}} = \curlyb{0, 1, 2, \hdots, N_{h}-1}$ and the values of $\overline{y}_{i,k}$ are drawn from the optimized state profile throughout the maneuver. One unique requirement of the FHMPC is that the number of steps in each horizon needs to be an odd number, since a horizon contains an integer number of control cycles. 
The evolution of the horizons in the FHMPC scheme is shown in \cref{fig:FHMPC_evolution_animation}. In this figure, the yellow area represents the initial optimized trajectory, which is computed only once over the entire horizon, while the fixed receding horizons are displayed in gray. In the figure, it is assumed that each horizon consists of $5$ steps, corresponding to $2$ control cycles.

It is important to note that the size of each horizon is fixed in terms of the number of steps, and not in terms of the actual duration of a horizon, aligning each of them with the fixed steps introduced by the first whole optimization. To keep the size of each horizon fixed, specifically those which approach the end of the maneuver, some artificial steps had to be introduced beyond the final time of the reconfiguration (refer to the last horizon in \cref{fig:FHMPC_evolution_animation}). The reference states for these artificial steps is drawn from propagating the linear dynamics in \cref{eq:ROE_dynamics_sol} with zero control input, and starting from $t_{f}$. The initial states are set to $\vec{y}_{i, m+1}$ for each satellite, which is the last step from the pre-optimized trajectory that was run over the whole maneuver. \\

The output of the control stepping function, whether the SHMPC or the FHMPC, consists of a set of acceleration vectors required to be executed by each deputy satellite’s onboard propulsion system. Notably, these acceleration vectors are provided by either MPC scheme in the RTN frame, necessitating an additional layer for each deputy. This layer translates the acceleration vector into a reference attitude for the attitude control system to track, as well as determines the required thrust level. Moreover, since the $L_{2}$ norm of the acceleration vector is allowed to be less the the minimum acceleration in the guidance layer as a result of the softening procedures, a saturation function is necessary to guarantee that the acceleration vector demanded by the control function is either bounded by the minimum and the maximum levels, or strictly zero. The adopted saturation scheme in this research is written as follows:
\begin{equation}
    \text{Sat}\parenth{\vec{u}, u_\text{min}, u_\text{max}} = \left\{ \begin{array}{ll}
        0, & \norm{\vec{u}} \leq \alpha u_\text{min}\\
        \dfrac{\vec{u}}{\norm{\vec{u}}} u_\text{min}, & \alpha u_\text{min} < \norm{\vec{u}} \leq u_\text{min} \\
        \vec{u}, & u_\text{min} < \norm{\vec{u}} \leq u_\text{max} \\
        \dfrac{\vec{u}}{\norm{\vec{u}}} u_\text{max}, & u_\text{max} < \norm{\vec{u}}\\
    \end{array}\right.
    \label{eq:saturation_function}
\end{equation}
where $0 \leq \alpha < 1$ is a user-defined parameter. In the context of the proposed guidance and control schemes, the last option of the saturation function (the fourth line) is never exploited since the guidance layer strictly constrains the $L_{2}$ norm of the acceleration vector to be less than or equal to the maximum allowable acceleration (see \cref{eq:umax_constraint_SOCP_soft}), unlike the soft constraints on the minimum allowable acceleration (see \cref{eq:umin_constraint_SOCP_soft}).\\

For the numerical simulations, the absolute position and velocity are propagated individually for each deputy, and the relative states are later extracted from the absolute ones. Moreover, surrogate models are used to account for estimation and pointing errors that affect the formation control system, as the development of these systems is beyond the scope of this work. A brief description of the adopted surrogate models is provided here for completeness.
In particular, the propagated absolute orbital elements are considered as the ground truth, with zero-mean normally distributed random noise added to generate estimated absolute states. The relative orbital elements are computed from the absolute elements of both the chief and each deputy, with the relative navigation outputs also generated by adding zero-mean normally distributed random noise. The variance-covariance matrices used to generate these noise signals are based on the absolute and relative navigation results from \cite{damico2010PhDThesis}.
Furthermore, attitude control errors are included in the simulations to model inaccuracies in the thruster firing direction. These errors are represented by introducing a disturbance in the satellite’s attitude, simulating a nominal pointing error based on data from the Triton-X data sheets.

\section{Results and discussion}\label{sec:Results}
In order to validate the proposed guidance and control schemes, they were run over the first and second formation reconfigurations in \ref{app:Reconfiguration_scenarios}. These two scenarios are specifically interesting because they form a basis to compare the performance of the closed-loop system with that of the open-loop system, presented in Tables \ref{tab:centralized_hard_soft_comparison_results_Tc_0.05} and \ref{tab:distributed_hard_soft_comparison_results_Tc_0.05}. The parameters used for the guidance scheme are the same as the ones in \cref{tab:guidance_soft_hard_comparison_parameters} except for the duration of the thrust arc, which is set to $T_{f}=0.05$ orbits. The parameters that relate to the SHMPC as well as the FHMPC are the sampling time, set to $50$ seconds for both MPC schemes, the horizon size, $N_{h}$, for the FHMPC case, which is fixed to $21$ steps, and the parameter $\alpha$, set to $40\%$ for both schemes.\\

Due to the inherent randomness in navigation and pointing errors, the two MPC strategies were evaluated over the two reconfiguration scenarios with $100$ repeated runs for both the centralized and the distributed settings. For each run, the total required Delta-V, the final state errors, and the maximum intrusion of one satellite to the KOZ of another, were recorded and then averaged across the $100$ trials. The results, presented in \cref{tab:Control_results_SHMPC_FHMPC}, summarize the total required Delta-V, the maximum and mean $L_{2}$ norm of the state error at the final time of the maneuver, and the maximum collision avoidance violation, for each MPC strategy.

{\renewcommand{\arraystretch}{\arraystretchfortable}
\begin{table*}[ht]
    \centering
    \caption{Results of the proposed control strategies over the two case studies}
    \label{tab:Control_results_SHMPC_FHMPC}
    \footnotesize
    \begin{tabular}{lccccccccccc}
    \hline
    \hline 
    ~ & \multicolumn{5}{c}{Centralized} & ~ & \multicolumn{5}{c}{Distributed} \\
    \cline{2-6} \cline{8-12}
    ~ & \multicolumn{2}{c}{Reconfiguration 1} & ~ & \multicolumn{2}{c}{Reconfiguration 2} & ~ & \multicolumn{2}{c}{Reconfiguration 1} & ~ & \multicolumn{2}{c}{Reconfiguration 2}\\
    \cline{2-3} \cline{5-6} \cline{8-9} \cline{11-12}
    ~ & SHMPC & FHMPC & ~ & SHMPC & FHMPC & ~ & SHMPC & FHMPC & ~ & SHMPC & FHMPC\\ 
    \hline 
    $\Delta V$ [m/s] & 2.78 & 2.69 & ~ & 1.67 & 1.58 & ~ & 3.02 & 2.77 & ~ & 1.65 & 1.58 \\ 
    $\text{max}\parenth{\norm{\vec{y}_{i,f} - \overline{\vec{y}}_{i,f}}}$ [m] & 2.03 & 1.57 & ~ & 2.40 & 1.37 & ~ & 7.21 & 1.49 & ~ & 2.19 & 0.93 \\ 
    $\text{mean}\parenth{\norm{\vec{y}_{i,f} - \overline{\vec{y}}_{i,f}}}$ [m] & 0.82 & 1.20 & ~ & 1.68 & 1.05 & ~ & 1.90 & 1.06 & ~ & 1.42 & 0.90 \\ 
    $\text{max}\parenth{R_{CA} - \norm{\vec{y}_{i,k} - \vec{y}_{j,k}}}$ [m] & 0.13 & 6.90 & ~ & 0.17 & 5.29 & ~ & 12.00 & 1.02 & ~ & 0.04 & 0.00 \\ 
    \hline 
    \hline 
    \end{tabular}
\end{table*}
}


The results in \cref{tab:Control_results_SHMPC_FHMPC} are best understood in comparison to the open-loop results of the softened problem, presented in Tables \ref{tab:centralized_hard_soft_comparison_results_Tc_0.05} and \ref{tab:distributed_hard_soft_comparison_results_Tc_0.05}. From a Delta-V perspective, both MPC schemes, SHMPC and FHMPC, demonstrate similar or slightly higher Delta-V requirements compared to the open-loop performance in both centralized and distributed settings. Notably, for these two specific reconfigurations, the SHMPC tends to demand slightly more Delta-V than the FHMPC, which exhibits strong tracking capabilities of the reference open-loop trajectory. However, as maneuver durations increase, the FHMPC begins to deviate more significantly from the reference trajectory, a trend that will be further explored later. These deviations can be attributed to longer maneuvers being more susceptible to discrepancies between the linear model used in the guidance strategy and the nonlinear model employed for state propagation in validation simulations. In particular, these deviations are evident in the convolution matrix presented in \cref{eq:ROE_dynamics_sol}, where certain assumptions inherent to its derivation play a significant role \cite{diMauro2018continuous}.\\

While the control accuracy of the SHMPC appears lower than that of the FHMPC for the specific reconfigurations examined here, this is not a general trend, as will be illustrated in subsequent discussions. The primary strength of the SHMPC lies in its adaptability. For example, it can dynamically improve performance when relative navigation accuracy is enhanced, such as during a rendezvous task when a vision-based navigation system is activated. Additionally, the SHMPC demonstrates superior resilience to elevated disturbance levels, including those arising from nonlinear dynamics not captured by the linear model. This adaptability is due to the SHMPC's approach of optimizing the entire trajectory from the current epoch to the end of the maneuver, making it particularly suitable for longer or disturbance-prone maneuvers. Conversely, the FHMPC optimizes over a shorter, fixed horizon and focuses on tracking a pre-optimized trajectory. This approach makes the FHMPC more vulnerable to control inaccuracies or higher Delta-V demands in scenarios involving prolonged maneuvers, increased navigation errors, or disturbance-heavy conditions.\\

Regarding collision risk, the FHMPC shows strong tracking performance in the distributed setting, leading to safe trajectories throughout the Monte Carlo simulations. However, in the centralized setting, the SHMPC clearly outperforms the FHMPC by producing collision-free relative trajectories, even in the presence of navigation and pointing errors. Violations of collision avoidance criteria can be traced to two main factors: the softening of collision avoidance constraints and the piecewise safety guaranteed by the guidance layer, which only ensures collision-free trajectories at specific sampling times. 
As indicated in \cref{tab:Control_results_SHMPC_FHMPC}, although the safety violations are generally within acceptable levels—given that $\beta_{max}$ was set to $10$ m in \cref{tab:guidance_soft_hard_comparison_parameters}—there may still be a desire to eliminate these violations entirely. This can be achieved either by adopting a smaller sampling time in the guidance layer, which would increase computational demand, or by introducing an artificial margin to the KOZ radius for each satellite to further enhance safety.\\

\begin{figure*}[ht]
\begin{subfigure}[c]{\columnwidth}
    \centering
    \includegraphics[width=\columnwidth]{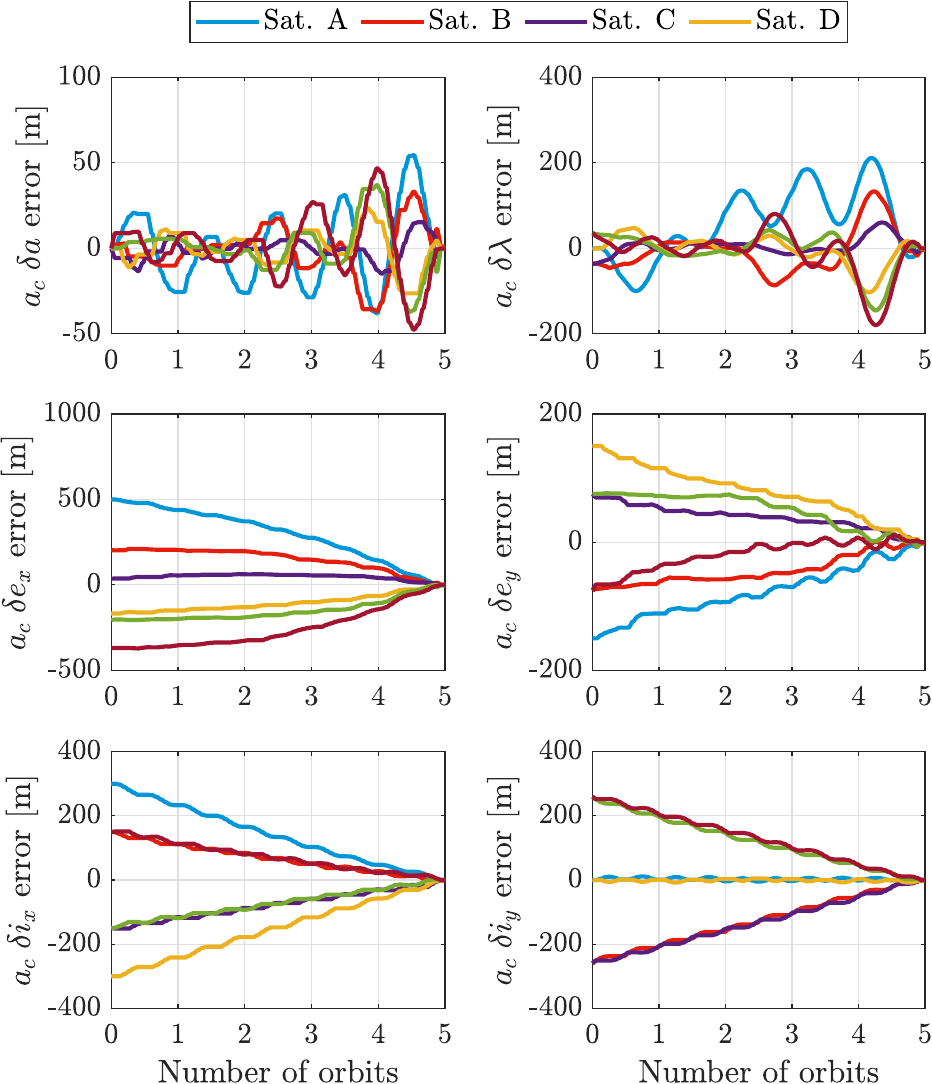}
    \caption{SHMPC} 
\end{subfigure}
\hfill
\begin{subfigure}[c]{\columnwidth}
    \centering
    \includegraphics[width=\columnwidth]{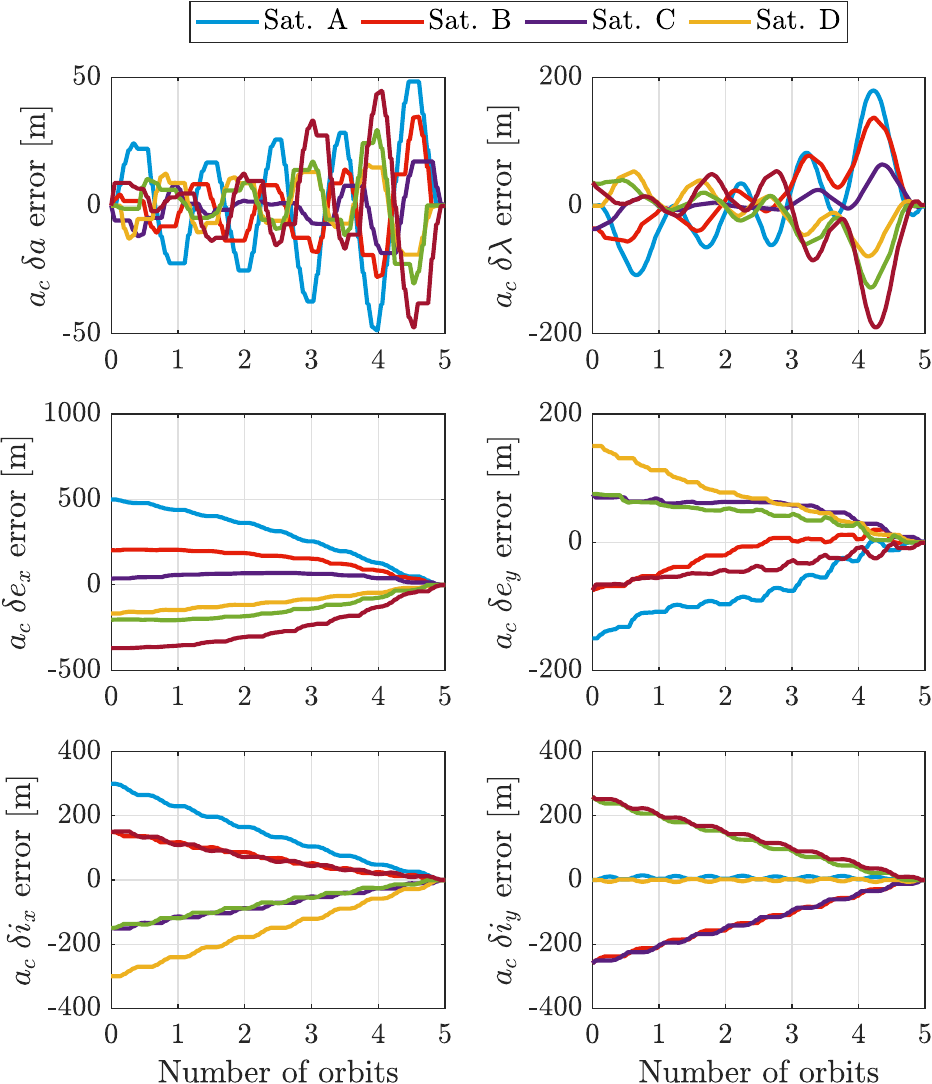}
    \caption{FHMPC}
\end{subfigure}
\caption{State error profiles for Reconfiguration 1 in the centralized setting.}
\label{fig:P1_ROE_error}
\end{figure*}

The state error profiles for the six satellites involved in the first reconfiguration scenario are presented in \cref{fig:P1_ROE_error}, comparing the SHMPC and FHMPC approaches for a single Monte Carlo run out of the $100$ conducted. For clarity and conciseness, only the results of the centralized approach are shown. It can be seen that each element of the dimensional ROE error vector gradually converges to zero by the end of the maneuver, achieving the primary control objective.

\begin{figure*}[ht]
\begin{subfigure}[c]{\columnwidth}
    \centering
    \includegraphics[width=\columnwidth]{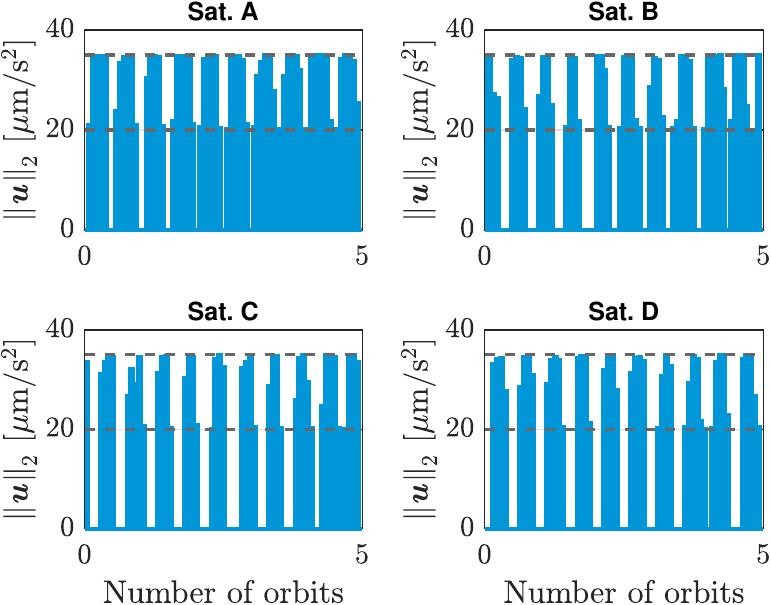}
    \caption{SHMPC} 
\end{subfigure}
\hfill
\begin{subfigure}[c]{\columnwidth}
    \centering
    \includegraphics[width=\columnwidth]{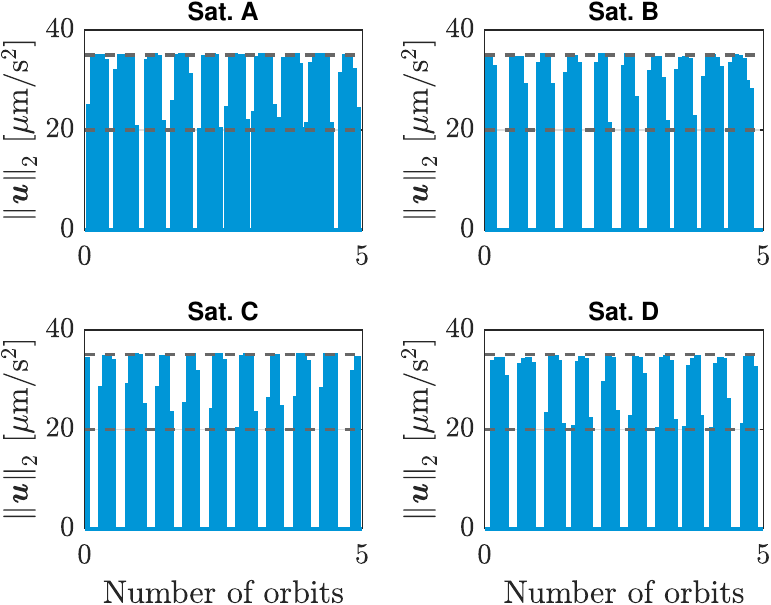}
    \caption{FHMPC}
\end{subfigure}
\caption{Control acceleration profiles for Reconfiguration 1 in the centralized setting.}
\label{fig:P1_acc_profile}
\end{figure*}

Beyond verifying the achievement of the primary control objective, it is also necessary to assess the adherence to specific constraints, such as the minimum acceleration threshold and collision avoidance. These aspects are investigated here. The control acceleration profiles corresponding to the same Monte Carlo run as in \cref{fig:P1_ROE_error} are shown in \cref{fig:P1_acc_profile}. The effect of the saturation function defined in \cref{eq:saturation_function} is evident, as the controller ensures that the satellite never exceeds feasible acceleration limits, even though the guidance layer allows slight violations of the minimum acceleration constraint.

\begin{figure*}[ht]
\begin{subfigure}[c]{\columnwidth}
    \centering
    \includegraphics[width=\columnwidth]{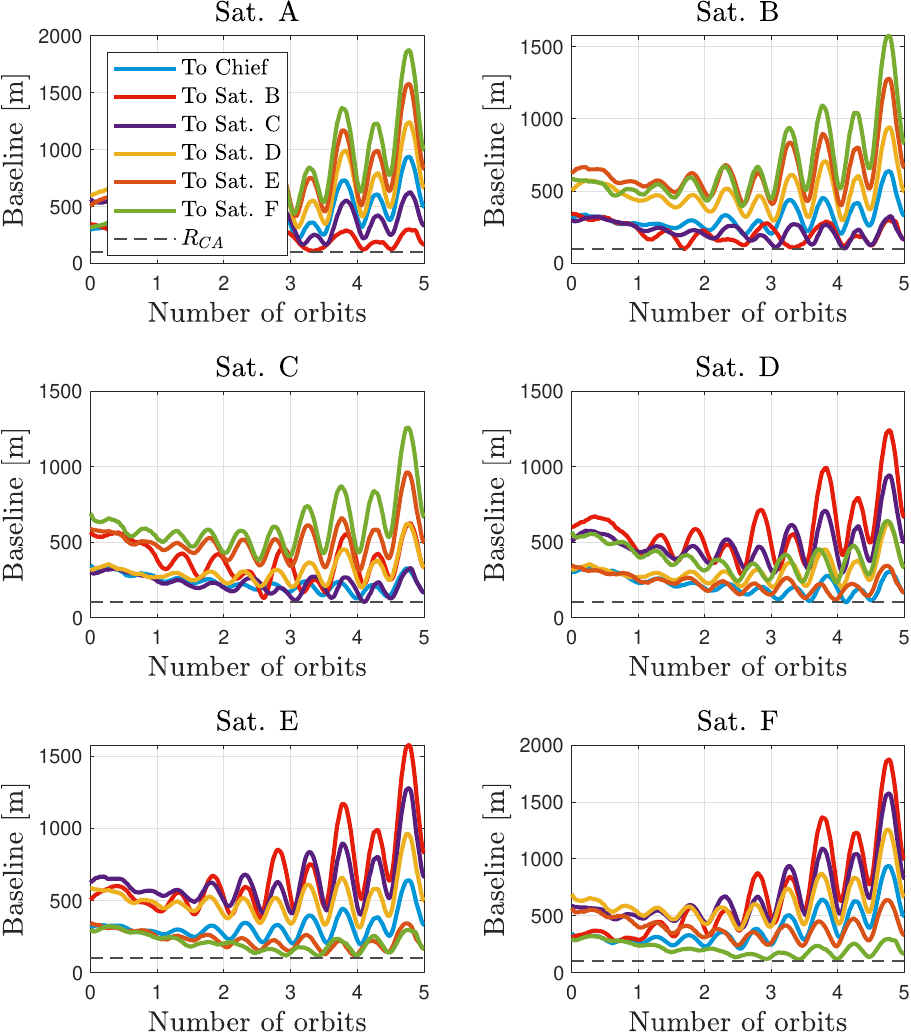}
    \caption{SHMPC} 
\end{subfigure}
\hfill
\begin{subfigure}[c]{\columnwidth}
    \centering
    \includegraphics[width=\columnwidth]{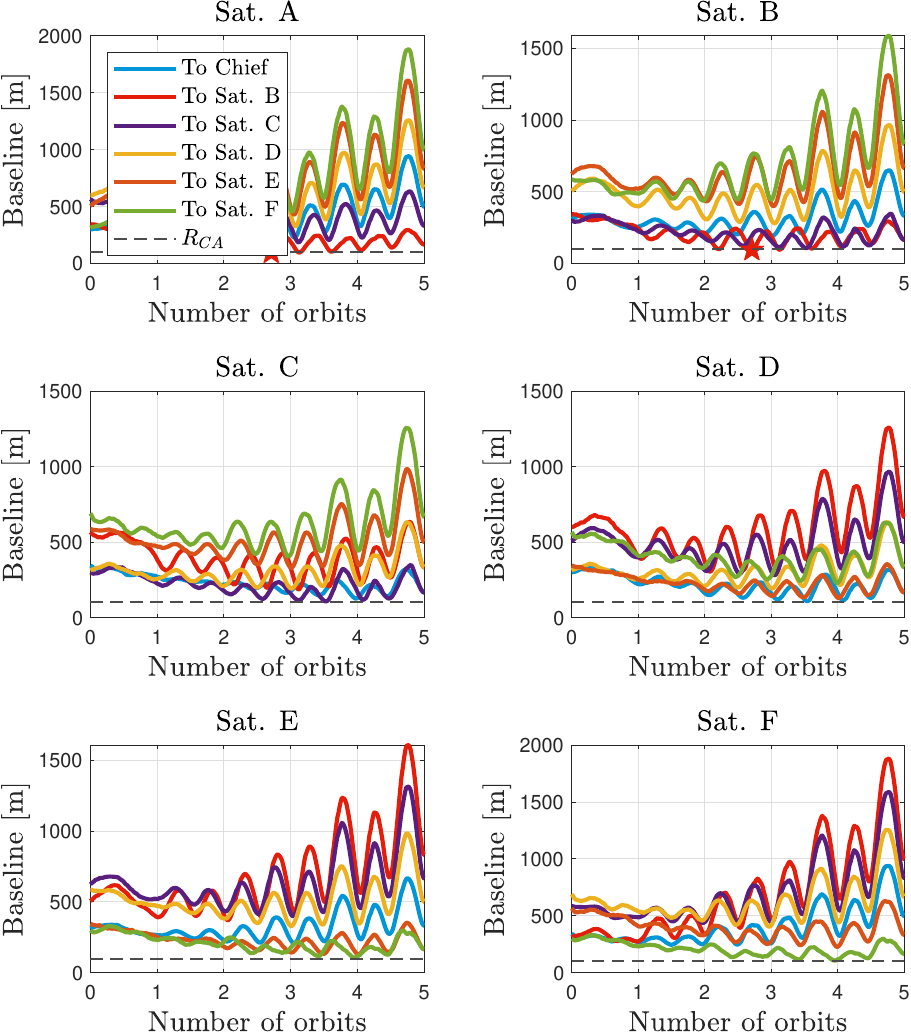}
    \caption{FHMPC}
\end{subfigure}
\caption{Baseline distances between satellites for Reconfiguration 1 in the centralized setting.}
\label{fig:P1_baseline}
\end{figure*}

Finally, the adherence to collision avoidance constraints is evaluated by examining the baselines between any two satellites throughout the maneuver. These baseline distances, corresponding to the same Monte Carlo run shown in \cref{fig:P1_ROE_error} and \cref{fig:P1_acc_profile}, are depicted in \cref{fig:P1_baseline}. A star symbol highlights instances of collision avoidance violations. For this particular Monte Carlo run, the FHMPC exhibits a $5$ m intrusion of satellite A into the KOZ of satellite B. This aligns with the results in \cref{tab:Control_results_SHMPC_FHMPC}, which indicate that, in the centralized setting, the FHMPC experiences larger violations of the collision avoidance threshold compared to the SHMPC. 
It is also worth noting that, as the results in \cref{fig:P1_baseline} correspond to the centralized approach, the baselines of each satellite relative to the chief spacecraft are included. This is because, in the centralized setting, the chief spacecraft is treated as a physical satellite.
 
The SHMPC and FHMPC schemes, in both their centralized and distributed configurations, were benchmarked against other MPC approaches from the literature. Specifically, the MPC algorithm presented in \cite{Belloni2023Relative} was selected as a reference for comparison. This benchmark experiment was conducted using Reconfiguration 3, detailed in \ref{app:Reconfiguration_scenarios}, which was initially introduced in \cite{Belloni2023Relative}. To ensure a fair comparison, both navigation and pointing errors were set to zero, consistent with the conditions in \cite{Belloni2023Relative}. \\

A summary of the results for the three MPC schemes is provided in \cref{tab:Benchmark_results}. The table includes the final state errors for each satellite, the mean final state error across all satellites ($\text{mean}\parenth{\norm{\vec{y}_{i,f} - \overline{\vec{y}}_{i,f}}}$), the Delta-V requirements for each satellite, and the total Delta-V for the maneuver. Additionally, the table highlights the improvements achieved by the SHMPC and FHMPC schemes relative to the reference MPC, expressed as a percentage reduction in the mean final state error and the total Delta-V. 
The benchmark experiment was conducted using the same parameters listed in \cref{tab:guidance_soft_hard_comparison_parameters}, except for the thrust arc duration, which was set to $T_{f} = 0.1$ orbits and the weighting matrices for the guidance problem which were adjusted to $\mat{Q} = 100 \cdot \mat{I}_{6}$ and $\mat{R} = 1.1 \cdot \mat{I}_{3}$. The MPC parameters used are the same as the ones in the earlier comparison experiments. It is noteworthy that all generated trajectories during this benchmark experiment were collision-free.
{\renewcommand{\arraystretch}{\arraystretchfortable}
\begin{table*}[ht]
    \centering
    \footnotesize
    \caption{Benchmark summary}
    \begin{tabular}{lccccccccccccc}
    \hline 
    \hline 
    ~ & \multicolumn{6}{c}{$L_{2}$ Norm of the final dimensional ROE error [m]} & ~ & \multicolumn{6}{c}{Required Delta-V [$\text{m}/\text{s}$]}\\
    \cline{2-7} \cline{9-14}
    ~ & Sat. A & Sat. B & Sat. C & Sat. D & Mean & Improved & ~ & Sat. A & Sat. B & Sat. C & Sat. D & Total & Saved \\ 
    \hline  
    Ref. MPC & 2.96 & 4.46 & 7.42 & 3.39 & 4.56 & 0\% & ~ & 0.09 & 0.97 & 0.91 & 0.12 & 2.09 & 0\% \\ 
    Cent. SHMPC & 1.08 & 0.13 & 0.11 & 1.08 & 0.60 & 86.81\% & ~ & 0.08 & 0.77 & 0.82 & 0.08 & 1.76 & 15.82\% \\ 
    Cent. FHMPC & 5.76 & 0.90 & 0.34 & 7.32 & 3.58 & 21.35\% & ~ & 0.27 & 0.74 & 0.77 & 0.24 & 2.02 & 3.30\% \\ 
    Dist. SHMPC & 1.07 & 0.12 & 0.10 & 1.08 & 0.59 & 86.98\% & ~ & 0.08 & 0.86 & 0.89 & 0.08 & 1.91 & 8.41\% \\ 
    Dist. FHMPC & 5.74 & 1.13 & 0.94 & 7.48 & 3.82 & 16.09\% & ~ & 0.27 & 0.79 & 0.82 & 0.24 & 2.12 & -1.26\% \\ 
    \hline 
    \hline 
    \end{tabular}
\label{tab:Benchmark_results}
\end{table*}
}

The results in \cref{tab:Benchmark_results} demonstrate that both SHMPC and FHMPC offer significant improvements over the reference MPC in terms of Delta-V efficiency and final state accuracy, particularly in the centralized setting. Despite the reference MPC using a much smaller sampling time for its guidance layer (50 seconds, compared to $0.1 \text{orbits} \approx 600$ seconds for SHMPC and FHMPC), which permits the reference MPC a much finer control authority, the SHMPC outperformed it significantly in terms of both control accuracy and Delta-V demand. These findings validate the suitability of SHMPC for longer maneuvers, as its optimization approach dynamically adapts to discrepancies in the system states.
The FHMPC also demonstrated measurable benefits over the reference MPC, particularly in the centralized configuration, where it achieved better final state accuracy and required less Delta-V. However, its performance ranked second to that of the SHMPC, which remains the most efficient strategy for the reconfiguration scenario considered in this study. 
Overall, the results confirm that adopting either SHMPC or FHMPC can significantly enhance satellite formation control performance compared to the reference MPC. Notably, SHMPC stands out as particularly advantageous for longer maneuvers, offering more precise control while demanding less fuel. Moreover, a comparison between the results of the centralized and distributed approaches reveals that the centralized setting consistently yields more fuel-efficient Delta-V trajectories, as is generally anticipated.\\

\begin{figure}[ht]
    \centering
    \includegraphics[width=\columnwidth]{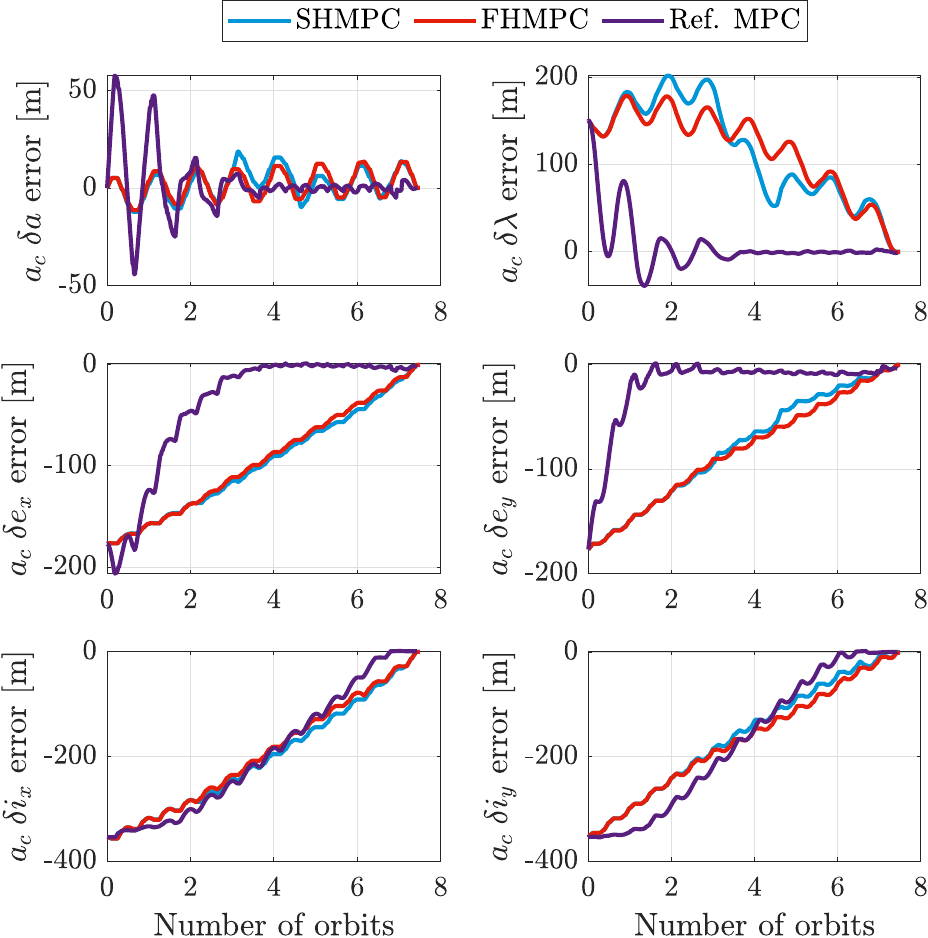}
    \caption{State error profiles of Sat. A in Reconfiguration 3 in the centralized setting}
    \label{fig:P5_2_ROE_error}
\end{figure}

The state error profiles for Sat. A across each of the three MPC schemes involved in the benchmark (in their centralized setting) are shown in \cref{fig:P5_2_ROE_error}.

The fact that the reference MPC does not rely on a pre-optimized trajectory enhances its resilience to unexpected external disturbances, similar to the SHMPC. However, the absence of a constraint on the maneuver's final time, combined with its objective of tracking the required final state at each MPC horizon makes it naturally inclined toward quicker maneuvers rather than fuel-efficient, longer-duration ones, as depicted in \cref{fig:P5_2_ROE_error}. This tendency to prioritize rapid convergence to the final state from the start of the maneuver makes the reference MPC more suitable for formation keeping rather than formation reconfiguration, as discussed in \cite{Mahfouz2023Autonomous}.
While the FHMPC closely resembles the reference MPC, its adherence to an optimized reference trajectory makes it more fuel-efficient under standard conditions. However, this tracking approach also makes the FHMPC more sensitive to external disturbances than the SHMPC or the reference MPC.\\

\begin{figure}[ht]
    \centering
    \includegraphics[width=\columnwidth]{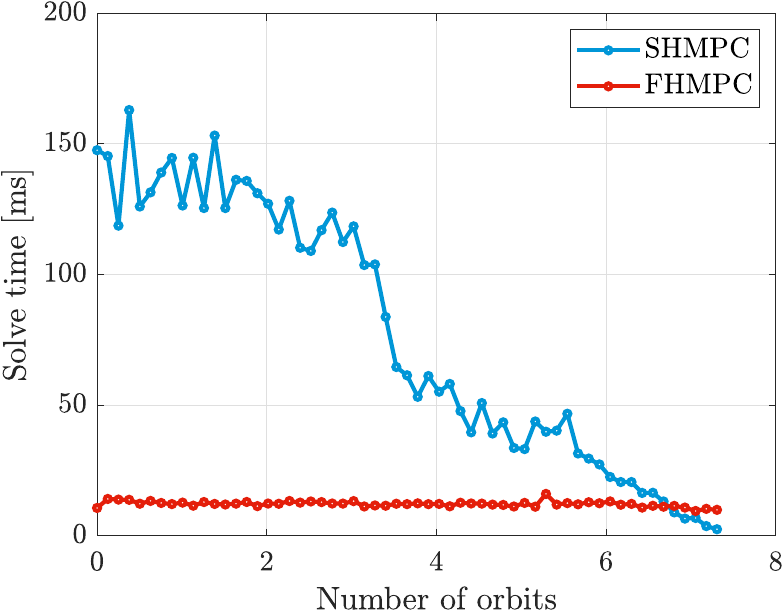}
    \caption{Solve times over Reconfiguration 3 in the centralized setting}
    \label{fig:P5_2_solve_time}
\end{figure}

The benchmark experiment clearly demonstrates that, typically, the SHMPC not only is able to achieve  more precise final states than the FHMPC, but also requires less Delta-V to complete the maneuver. However, this comes at the expense of increased computational complexity. Unlike the FHMPC, which maintains a fixed optimization horizon, the SHMPC utilizes a variable horizon that starts at a maximum length at the beginning of the maneuver and gradually decreases toward the end. A larger horizon length requires more time to solve its associated guidance problem. This is illustrated in \cref{fig:P5_2_solve_time}, where the computation time for solving the guidance problem at each optimization horizon is shown for both the SHMPC and FHMPC.
The absolute values presented in \cref{fig:P5_2_solve_time} are not crucial for our analysis, as they can vary depending on the onboard processor, the solver used, and other implementation factors such as the programming language. Instead, the focus is on the trend of each line. The solve time for each FHMPC horizon remains nearly constant, reflecting its fixed horizon length, whereas the solve time for each SHMPC horizon decreases as the horizon length reduces. For longer maneuvers, the SHMPC may become considerably more computationally demanding.

The advantages of the FHMPC extend beyond its stable and predictable computational complexity as it is also well-suited for onboard implementation. By keeping the horizon length fixed, the size of the recurrent guidance problem is also fixed, which simplifies the writing/generation of embedded software without the need for dynamic memory allocation. This characteristic is especially beneficial in developing safety-critical software, such as that used onboard spacecraft, where dynamic memory allocation is often avoided to enhance reliability and safety \cite{Holzmann2006NASA_10_rules}. Indeed, SHMPC can also be implemented using static memory allocation, however, this process is significantly more challenging and requires certain programmatic techniques to manage the variable horizon effectively.

\section{Conclusion}\label{sec:Conclusion}
This study introduces two Model Predictive Control (MPC) schemes, Shrinking-Horizon MPC (SHMPC) and Fixed-Horizon MPC (FHMPC), tailored for satellite formation reconfiguration tasks. Each scheme was designed with considerations for both final state accuracy and fuel efficiency, while also addressing critical operational constraints such as maximum and minimum acceleration limits and collision avoidance.\\

To ensure feasible solutions under varying operational conditions, the proposed schemes incorporate a softening mechanism in the guidance layer. This approach is specifically valuable for managing cases where optimization horizons may initially appear infeasible. By softening constraints that are likely to induce infeasibility, namely, the final state, minimum acceleration, and collision avoidance constraints, the guidance layer enables the MPC to converge on a solution even under challenging conditions. The softening process introduces slack variables to relax the strict constraints, thus ensuring that the problem remains solvable by allowing minimal, controlled violations. This approach maintains constraint integrity while providing flexibility, and is especially useful in cases of tight reconfiguration timelines or significant environmental disturbances.\\

Based on the benchmark scenarios, the FHMPC is primarily recommended for short-duration maneuvers, while for extended reconfiguration scenarios, the SHMPC generally offers better performance, with enhanced accuracy and reduced Delta-V requirements compared to the FHMPC. However, these advantages come with the trade-off of increased computational complexity, as the SHMPC’s variable optimization horizon dynamically adjusts throughout the maneuver. This adaptability allows the SHMPC to manage unexpected disturbances more effectively but also increases computational demands, making it challenging for onboard implementation in reliability-critical systems, where dynamic memory allocation is to be avoided.
In contrast, the FHMPC operates with a fixed optimization horizon, ensuring stable and predictable computational loads. Although it generally requires slightly higher Delta-V than the SHMPC for long maneuvers, it still demonstrates notable improvements over reference MPC schemes from the literature.\\

The two MPC schemes are implemented in both centralized and distributed settings. While the centralized approach is generally more fuel-efficient, the distributed approach offers scalability for formations with a large number of satellites.\\

\section*{Acknowledgments}
This research was funded in whole, or in part, by the Luxembourg National Research Fund (FNR), grant reference BRIDGES/19/MS/14302465. For the purpose of open access, and in fulfilment of the obligations arising from the grant agreement, the author has applied a Creative Commons Attribution 4.0 International (CC BY 4.0) license to any Author Accepted Manuscript version arising from this submission.\\

\appendix
\section{Reconfiguration scenarios used in the verification experiments}
\setcounter{table}{0}
\setcounter{figure}{0}
\label{app:Reconfiguration_scenarios}
This appendix presents the initial and final states for the reconfiguration scenarios which are used in the verification experiments throughout the paper.  
Four reconfiguration scenarios have been identified for the benchmark experiments.

\subsection{Reconfiguration 1 - Multi-PCO to Multi-cartwheel}
In this reconfiguration scenario, 6 deputy satellites are assumed to be in a multi Projected Circular Orbit (PCO) configuration at $t_{0}$ and are required to be reconfigured into a multi-cartwheel at $t_{f}$, where $t_{f}-t_{0} = 5\; \text{orbits}$. \cref{tab:R1_init_final_states} summarizes the initial and final dimensional ROE vectors, $\vec{y}_{0}$ and  $\vec{y}_{f}$, in meters for all the deputies. The orbit of the chief is parameterized by the osculating orbital elements $\tilde{\vec{\alpha}}_{c}\parenth{t_{0}} = \begin{bmatrix} 6978 \;\text{km} & 90^{\circ} &  10^{-3} & 0 & 97.87^{\circ} & 0^{\circ} \end{bmatrix}^{\intercal}$ at the initial time of the maneuver.
{\renewcommand{\arraystretch}{\arraystretchfortable}
\begin{table*}[ht]
    \centering
    \caption{Initial and final (required) states for each of the deputies in Reconfiguration 1}
    
    \begin{tabular}{lcccccc}
        \hline
        \hline
        ~ & Sat. A & Sat. B & Sat. C & Sat. D & Sat. E & Sat. F \\
        \hline
        \rule{0pt}{\tablerowsep}
        $\vec{y}_{0}\; [\text{m}]$ & 
        $\begin{bmatrix} 0\\ 0\\ 0\\ -150\\ 300\\ 0 \end{bmatrix}$ & 
        $\begin{bmatrix} 0\\ -35.91\\ -129.90\\ -75\\ 150\\ -259.81 \end{bmatrix}$ & 
        $\begin{bmatrix} 0\\ -35.91\\ -129.90\\ 75\\ -150\\ -259.81 \end{bmatrix}$ & 
        $\begin{bmatrix} 0\\ 0\\ 0\\ 150\\ -300\\ 0 \end{bmatrix}$ &
        $\begin{bmatrix} 0\\ 35.91\\ 129.90\\ 75\\ -150\\ 259.81 \end{bmatrix}$ &
        $\begin{bmatrix} 0\\ 35.91\\ 129.90\\ -75\\ 150\\ 259.81 \end{bmatrix}$\\
        \rule{0pt}{\tablerowsep}
        $\vec{y}_{f}\; [\text{m}]$ & 
        $\begin{bmatrix} 0\\ 0\\ -500\\ 0\\ 0\\ 0 \end{bmatrix}$ & 
        $\begin{bmatrix} 0\\ 0\\ -333.33\\ 0\\ 0\\ 0 \end{bmatrix}$ & 
        $\begin{bmatrix} 0\\ 0\\ -166.67\\ 0\\ 0\\ 0 \end{bmatrix}$ & 
        $\begin{bmatrix} 0\\ 0\\ 166.67\\ 0\\ 0\\ 0 \end{bmatrix}$ &
        $\begin{bmatrix} 0\\ 0\\ 333.33\\ 0\\ 0\\ 0 \end{bmatrix}$ & 
        $\begin{bmatrix} 0\\ 0\\ 500\\ 0\\ 0\\ 0 \end{bmatrix}$\\
        \hline
        \hline
    \end{tabular}
    \label{tab:R1_init_final_states}
\end{table*}
}

\begin{figure*}[ht]
\begin{subfigure}[c]{\columnwidth}
    \centering
    \includegraphics[width=\columnwidth]{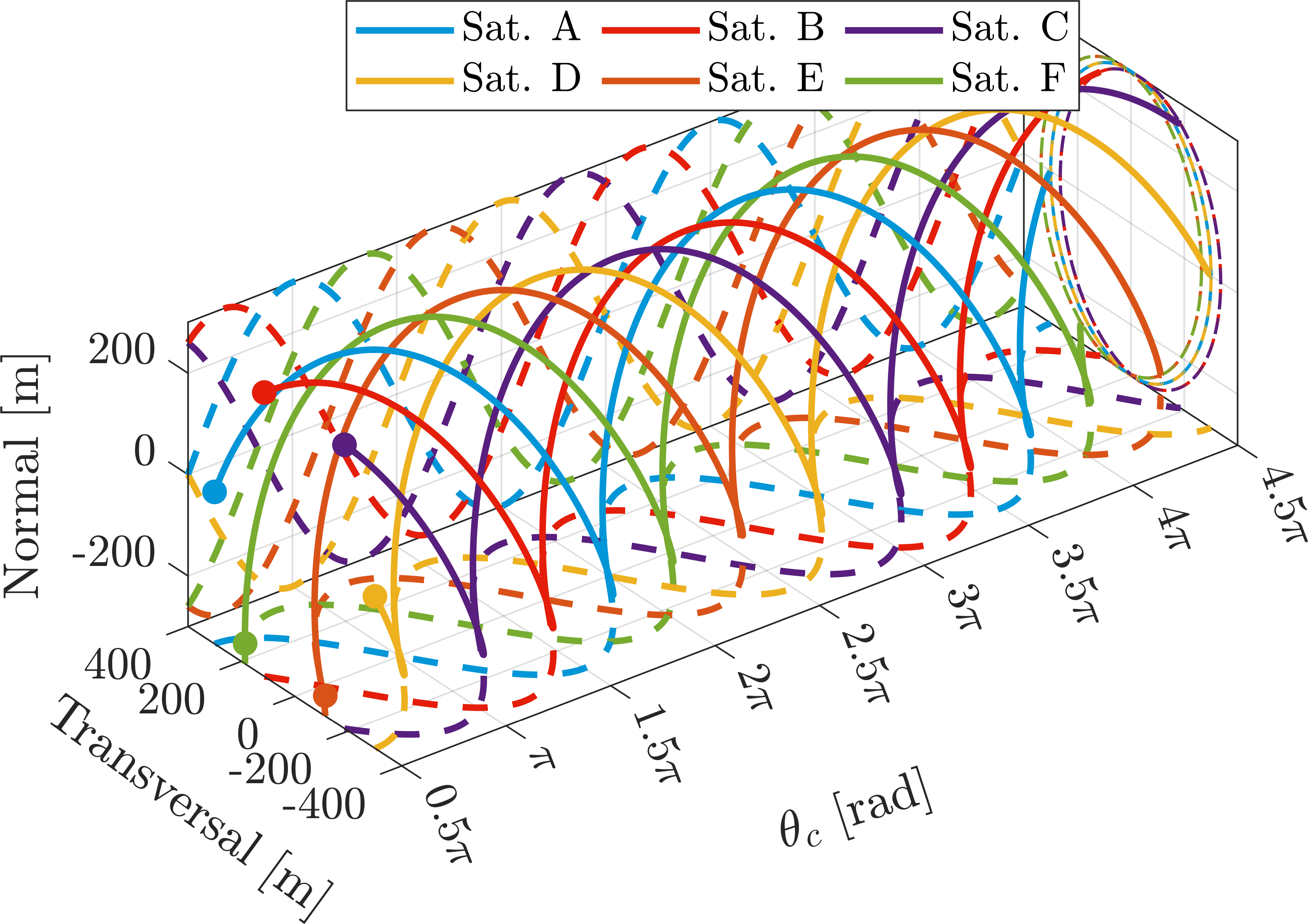}
    \caption{Initial relative orbits (Multi-PCO)} 
\end{subfigure}
\hfill
\begin{subfigure}[c]{\columnwidth}
    \centering
    \includegraphics[width=\columnwidth]{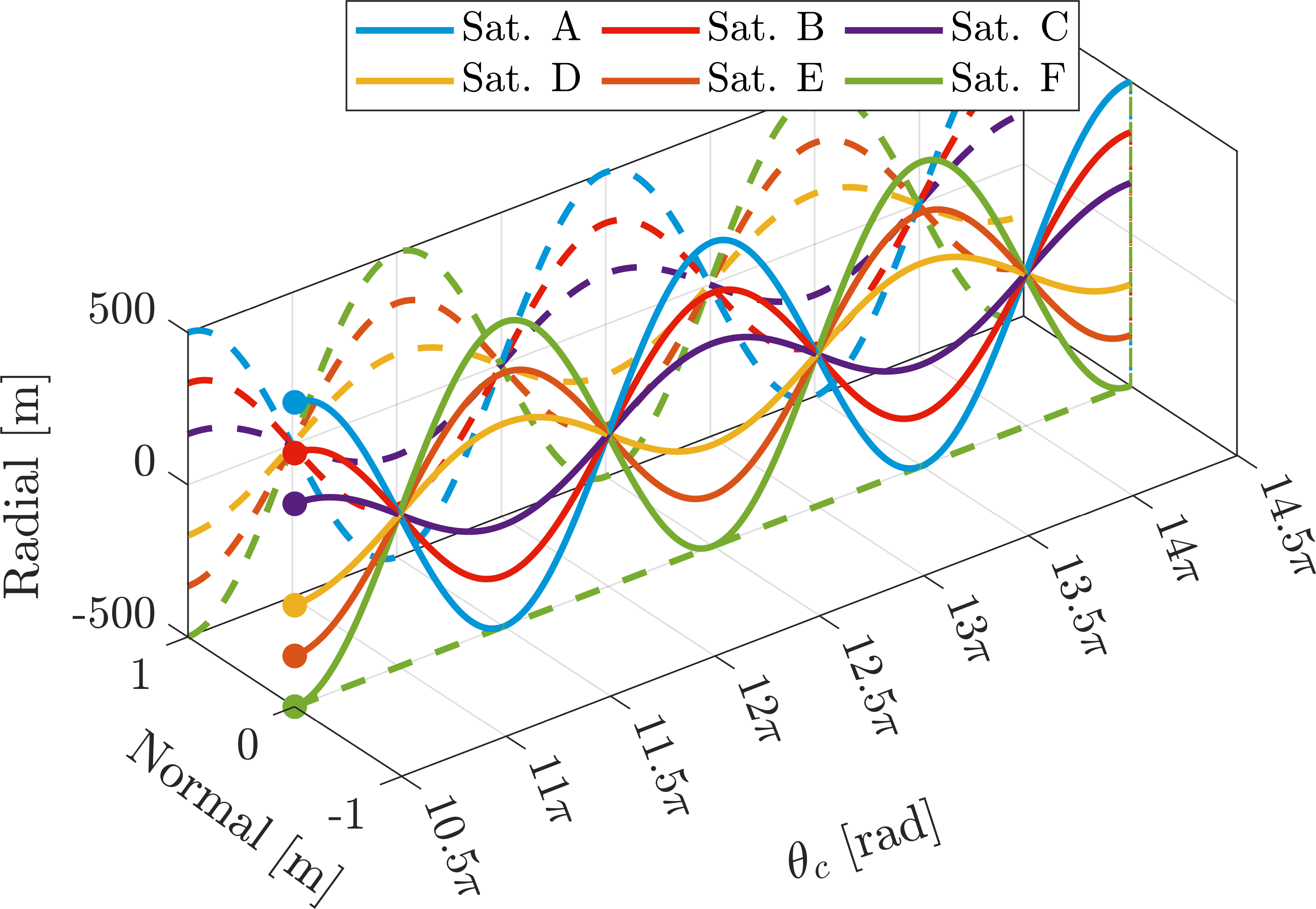}
    \caption{Final relative orbits (Multi-cartwheel)}
\end{subfigure}
\caption{Shapes of the unforced relative orbits involved in Reconfiguration 1}
\label{fig:P1_init_and_final_orbits}
\end{figure*}

The initial and final relative orbits are depicted in \cref{fig:P1_init_and_final_orbits}. This figure illustrates the relative trajectories the spacecraft would follow if no thrusting or disturbing forces are applied to them. 

\subsection{Reconfiguration 2 - Multi-cartwheel to Multi-helix}
In this reconfiguration scenario, 4 deputy satellites are assumed to be in a multi-cartwheel configuration at $t_{0}$ and are required to be reconfigured into a multi-helix configuration at $t_{f}$, where $t_{f}-t_{0} = 5\; \text{orbits}$. \cref{tab:R2_init_final_states} summarizes the initial and final dimensional ROE vectors, $\vec{y}_{0}$ and  $\vec{y}_{f}$, in meters for all the deputies. The orbit of the chief is parameterized by the osculating orbital elements $\tilde{\vec{\alpha}}_{c}\parenth{t_{0}} = \begin{bmatrix} 6978 \;\text{km} & 90^{\circ} &  10^{-3} & 0 & 97.87^{\circ} & 0^{\circ} \end{bmatrix}^{\intercal}$ at the initial time of the maneuver.
{\renewcommand{\arraystretch}{\arraystretchfortable}
\begin{table}[ht]
    \centering
    \caption{Initial and final (required) states for each of the deputies in Reconfiguration 2}
    \begin{tabular}{lcccc}
    \hline
    \hline
    ~ & Sat. A & Sat. B & Sat. C & Sat. D \\
    \hline
    \rule{0pt}{\tablerowsep}
    $\vec{y}_{0}\; [\text{m}]$ & 
    $\begin{bmatrix} 0 \\ 0 \\ -500 \\ 0 \\ 0 \\ 0 \end{bmatrix}$ & 
    $\begin{bmatrix} 0 \\ 0 \\ -250 \\ 0 \\ 0 \\ 0 \end{bmatrix}$ & 
    $\begin{bmatrix} 0 \\ 0 \\ 250 \\ 0 \\ 0 \\ 0 \end{bmatrix}$ & 
    $\begin{bmatrix} 0 \\ 0 \\ 500 \\ 0 \\ 0 \\ 0 \end{bmatrix}$ \\
    \rule{0pt}{\tablerowsep}
    $\vec{y}_{f}\; [\text{m}]$ & 
    $\begin{bmatrix} 0 \\ 34.56 \\ 0 \\ -250 \\ 0 \\ -250 \end{bmatrix}$ & 
    $\begin{bmatrix} 0 \\ 17.28 \\ 0 \\ -125 \\ 0 \\ -125 \end{bmatrix}$ & 
    $\begin{bmatrix} 0 \\ -17.28 \\ 0 \\ 125 \\ 0 \\ 125 \end{bmatrix}$ & 
    $\begin{bmatrix} 0 \\ -34.56 \\ 0 \\ 250 \\ 0 \\ 250 \end{bmatrix}$ \\
    \hline
    \hline
    \end{tabular}
    \label{tab:R2_init_final_states}
\end{table}
}

\begin{figure*}[ht]
\begin{subfigure}[c]{\columnwidth}
    \centering
    \includegraphics[width=\columnwidth]{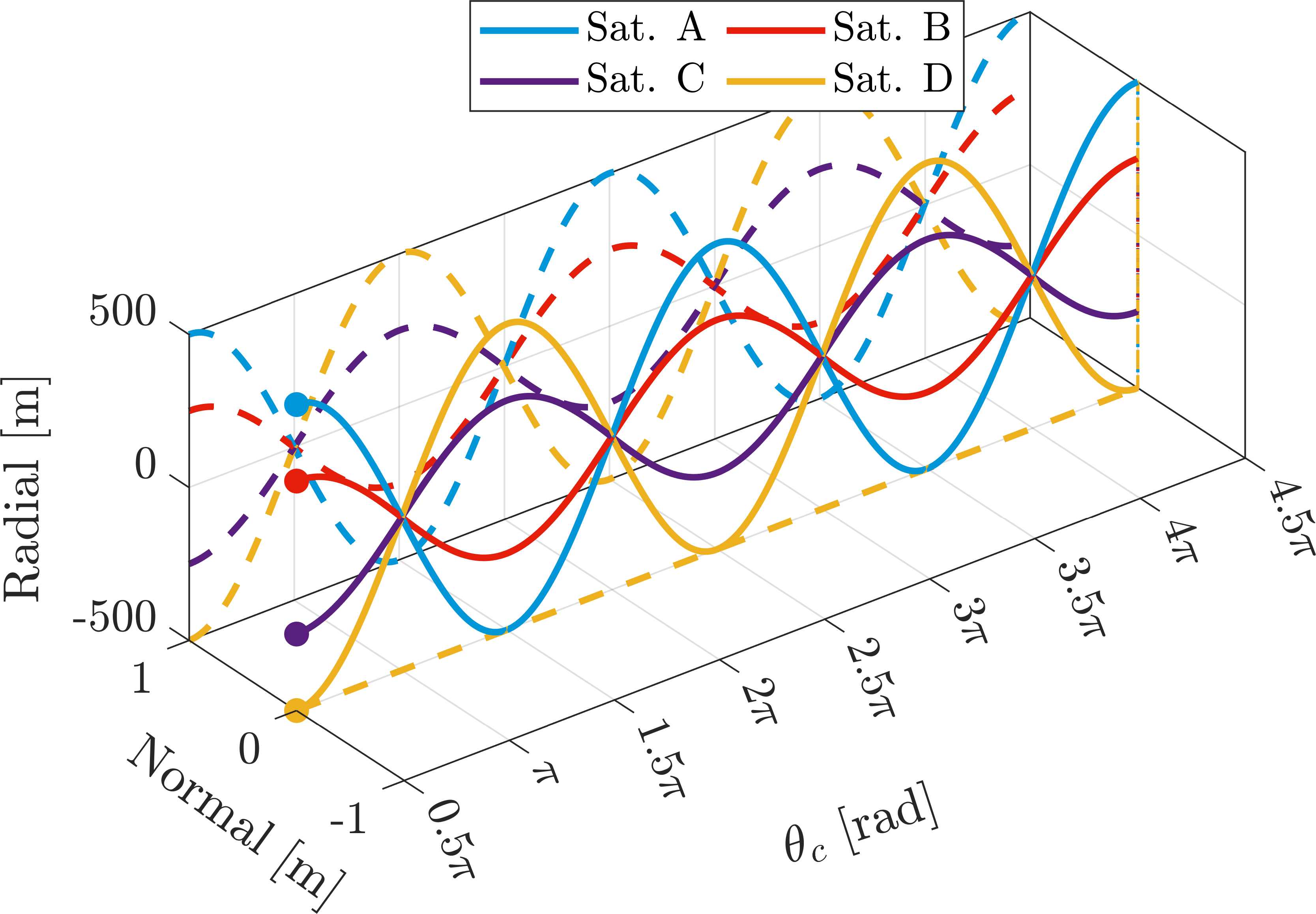}
    \caption{Initial relative orbits (Multi-cartwheel)} 
\end{subfigure}
\hfill
\begin{subfigure}[c]{\columnwidth}
    \centering
    \includegraphics[width=\columnwidth]{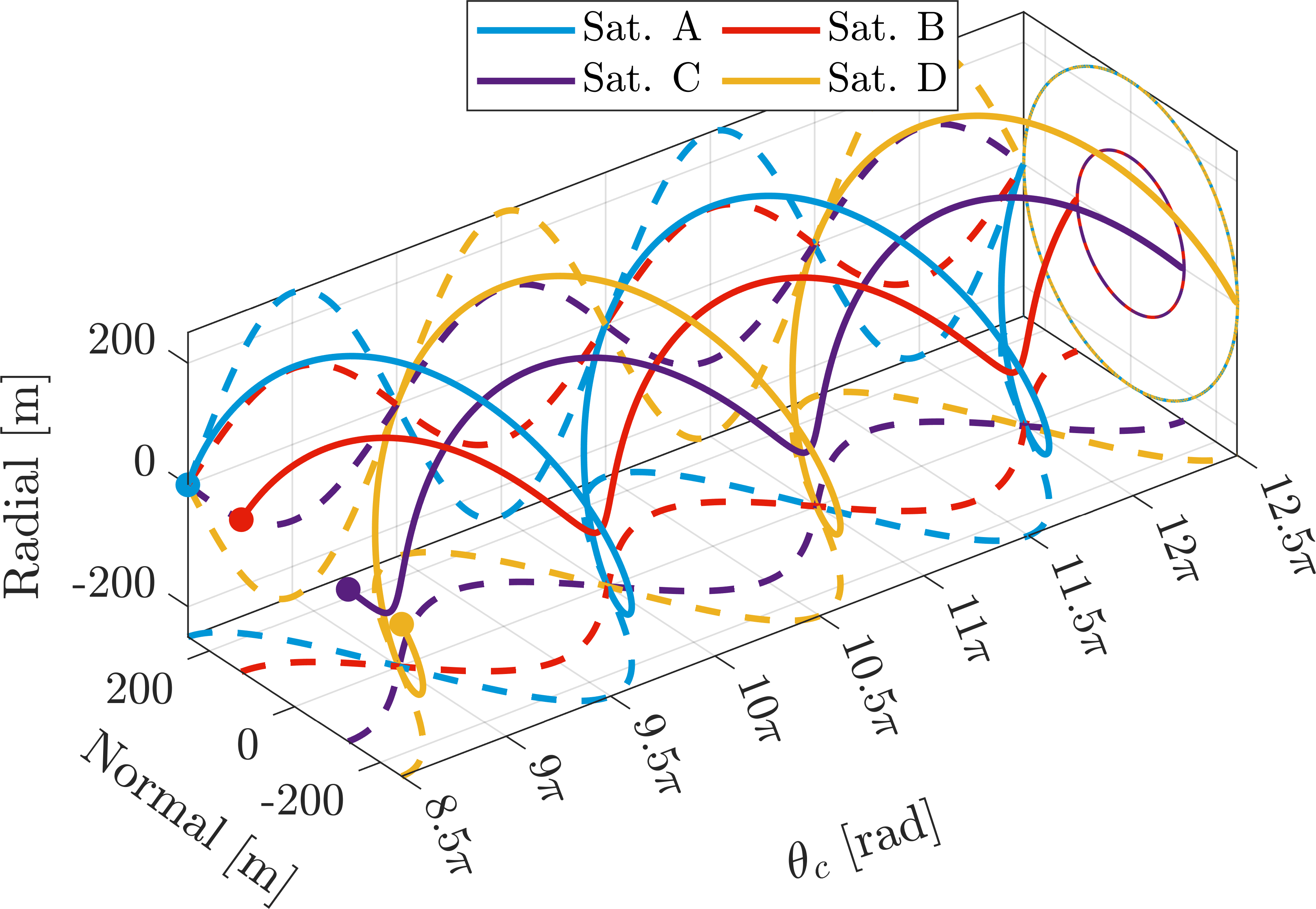}
    \caption{Final relative orbits (Multi-helix)}
\end{subfigure}
\caption{Shapes of the unforced relative orbits involved in Reconfiguration 2}
\label{fig:P2_init_and_final_orbits}
\end{figure*}

The unforced initial and final relative orbits of Reconfiguration 2 are shown in \cref{fig:P2_init_and_final_orbits}.

\subsection{Reconfiguration 3 - Trailing to Tetrahedron}
In this reconfiguration scenario, 4 deputy satellites are assumed to be in a trailing/coplanar configuration at $t_{0}$ and are required to be reconfigured into a tetrahedron shape at $t_{f}$, where $t_{f}-t_{0} = 7.5\; \text{orbits}$. \cref{tab:R3_init_final_states} summarizes the initial and final dimensional ROE vectors, $\vec{y}_{0}$ and  $\vec{y}_{f}$, in meters for all the deputies. The orbit of the chief is parameterized by the osculating orbital elements $\tilde{\vec{\alpha}}_{c}\parenth{t_{0}} = \begin{bmatrix} 6780.678 \;\text{km} & 90^{\circ} &  0 & 029 & 97^{\circ} & 30^{\circ} \end{bmatrix}^{\intercal}$ at the initial time of the maneuver.
{\renewcommand{\arraystretch}{\arraystretchfortable}
\begin{table}[ht]
    \centering
    \caption{Initial and final (required) states for each of the deputies in Reconfiguration 3}
    \begin{tabular}{ccccc}
    \hline
    \hline
    ~ & Sat. A & Sat. B & Sat. C & Sat. D \\
    \hline
    \rule{0pt}{\tablerowsep}
    $\vec{y}_{0}\; [\text{m}]$ & 
    $\begin{bmatrix} 0 \\ 750 \\ 0 \\ 0 \\ 0 \\ 0 \end{bmatrix}$ & 
    $\begin{bmatrix} 0 \\ 250 \\ 0 \\ 0 \\ 0 \\ 0 \end{bmatrix}$ & 
    $\begin{bmatrix} 0 \\ -250 \\ 0 \\ 0 \\ 0 \\ 0 \end{bmatrix}$ & 
    $\begin{bmatrix} 0 \\ -750 \\ 0 \\ 0 \\ 0 \\ 0 \end{bmatrix}$ \\
    \rule{0pt}{\tablerowsep}
    $\vec{y}_{f}\; [\text{m}]$ & 
    $\begin{bmatrix} 0 \\ 400 \\ 0 \\ 0 \\ 0 \\ 0 \end{bmatrix}$ & 
    $\begin{bmatrix} 0 \\ 100 \\ 177 \\ 177 \\ 354 \\ 354 \end{bmatrix}$ & 
    $\begin{bmatrix} 0 \\ -100 \\ -177 \\ 177 \\ -354 \\ 354 \end{bmatrix}$ & 
    $\begin{bmatrix} 0 \\ -400 \\ 0 \\ 0 \\ 0 \\ 0 \end{bmatrix}$ \\
    \hline
    \hline
    \end{tabular}
    \label{tab:R3_init_final_states}
\end{table}
}

\begin{figure*}[ht]
\begin{subfigure}[c]{\columnwidth}
    \centering
    \includegraphics[width=\columnwidth]{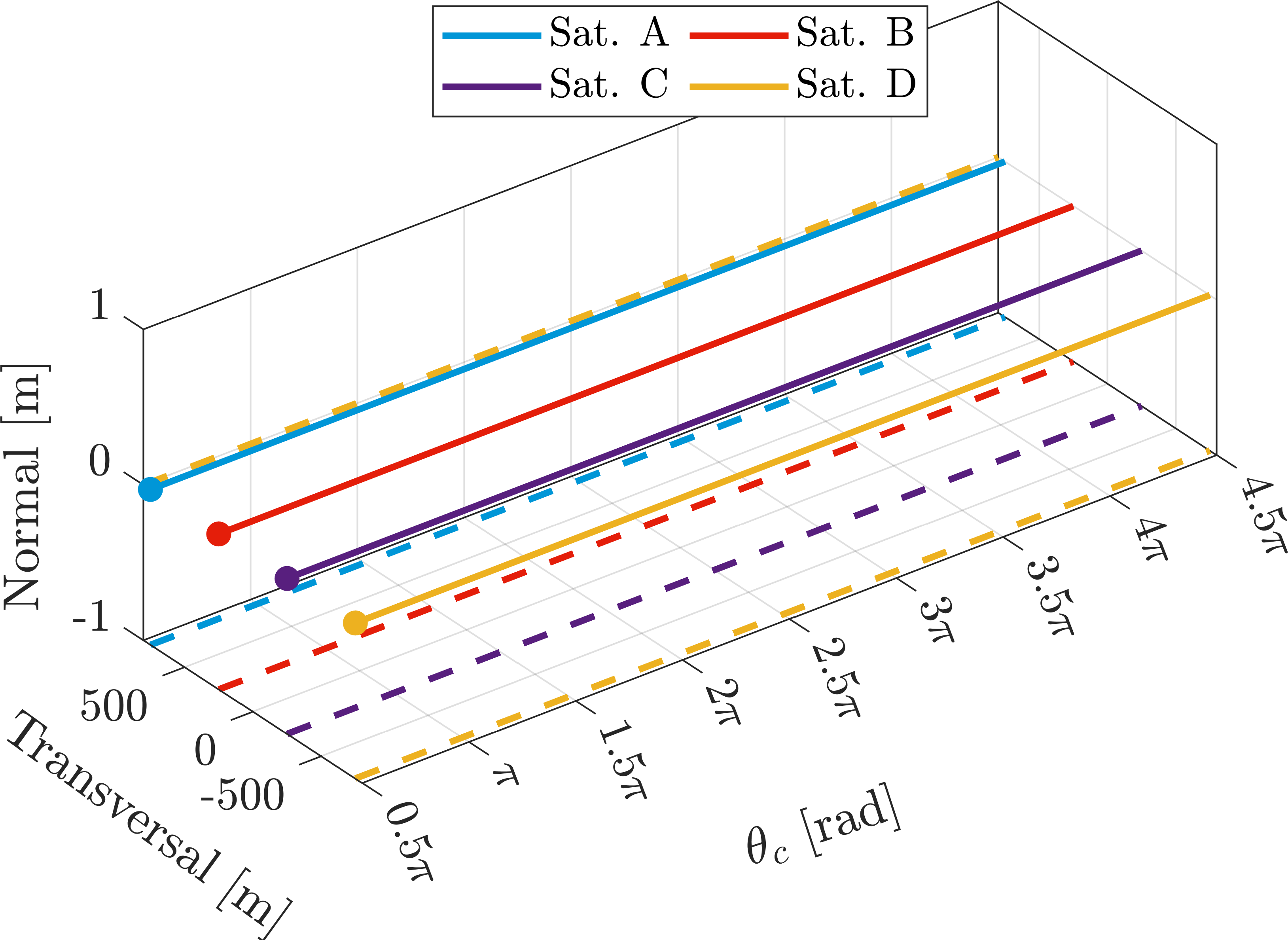}
    \caption{Initial relative orbits (Trailing)} 
\end{subfigure}
\hfill
\begin{subfigure}[c]{\columnwidth}
    \centering
    \includegraphics[width=\columnwidth]{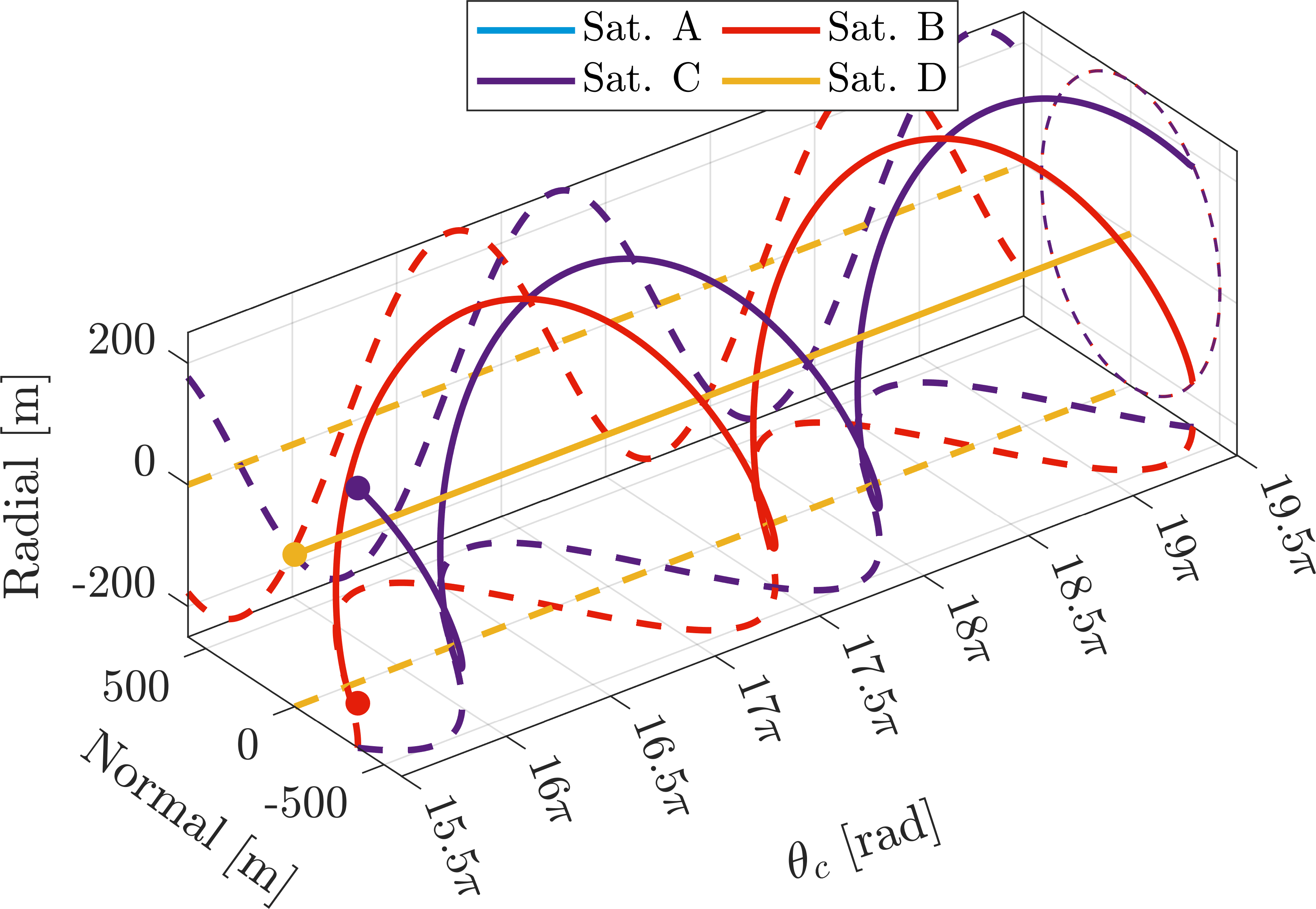}
    \caption{Final relative orbits (Tetrahedron)}
\end{subfigure}
\caption{Shapes of the unforced relative orbits involved in Reconfiguration 3}
\label{fig:P3_init_and_final_orbits}
\end{figure*}

The unforced initial and final relative orbits of Reconfiguration 2 are shown in \cref{fig:P3_init_and_final_orbits}.

Although Satellites A and D in the final configuration of Reconfiguration 3 seem to have the same relative orbit when viewed from the Normal-Radial plane, they actually follow two distinct orbits. This difference arises from the separation in the Transversal direction, which is reflected in their different mean arguments of longitude, as shown in \cref{tab:R3_init_final_states}. In fact, the tetrahedron configuration is a composed of two satellites flying in a helix configuration (Satellites B and C) and two others in a trailing formation (Satellites A and D).

 \bibliographystyle{elsarticle-num} 
 \bibliography{References.bib}
\end{document}